\documentclass[article,twocolumn,floatfix]{revtex4}

\usepackage{graphicx} 
\usepackage{subfig}
\usepackage{caption}
\usepackage{array}
\usepackage{color}

\newcommand{\ket}[1]{\left| #1 \right\rangle}
\newcommand{\bra}[1]{\left\langle #1 \right|}

\newcommand{\Tr}{\mathrm{Tr}}

\newcommand{\be}{\begin{equation}}
\newcommand{\ee}{\end{equation}}
\newcommand{\bea}{\begin{eqnarray}}
\newcommand{\eea}{\end{eqnarray}}

\usepackage{dcolumn}
\usepackage{bm}
\usepackage{amssymb,amsmath}
\usepackage{color}
\usepackage{float}

\definecolor{DarkGreen}{rgb}{0,0.6,0.2}

\begin{document}
\title{Single-photon time-dependent spectra in coupled cavity arrays}
\author{Imran M. Mirza$^1$, S.J. van Enk$^1$, H.J. Kimble$^2$}
\affiliation{$^1$Oregon Center for Optics and Department of Physics\\University of Oregon\\
Eugene, OR 97401}
\affiliation{$^2$California Institute of Technology 12-33\\
Pasadena CA 91125}
\begin{abstract}
We show how the input-output formalism for cascaded quantum systems combined with the quantum trajectory approach yields a compact and physically intuitive description of single photons propagating through a coupled cavity array. As a new application we obtain the time-dependent spectrum of such a single photon, which directly reflects the
fact that only certain frequency components of single-photon wavepackets are trapped inside the cavities and hence are delayed in time. We include in our description the actual generation of the single photon, by assuming we have a single emitter in one of the resonators.
\end{abstract}
  \maketitle
\section{Introduction}
There has been a substantial effort in the last decade or so
to find means of storing and/or delaying light by making use of arrays of coupled microcavities \cite{yariv1999coupled,heebner2002scissor,heebner2002slow,scheuer2005coupled,baba2008slow,krauss2008we}.
The main effort has been aimed at storing and delaying classical light pulses for classical communication purposes, but in principle the same ideas extend down to the single-photon level (for theory, see \cite{shen2005coherent,shen2009theory1,shen2009theory,rephaeli2012stimulated}), so that quantum communication protocols may benefit as well from these efforts. 

We focus here on the single-photon case. Single photons are not so easy to produce, but one possible technique of producing single photons on demand fits very well with the system under consideration here, namely, to use a single emitter (an atom or a quantum dot or an NV center in diamond) inside a cavity \cite{santori2002indistinguishable,kuhn2002deterministic,mckeever2004deterministic,englund2010,riedrich2011}. We include the production of the single photon in our description by assuming we have a single atom inside one of the resonators.  We study then the photon's properties when it is propagating through one or more additional (empty) cavities. This study will thereby be relevant to the development of deterministic single photon sources, for, e.g., quantum key distribution.

We will describe our quantum system by means of the input-output formalism  for cascaded systems \cite{carmichael1993quantum,gardiner1993driving}. This formalism is eminently suited for describing coupled cavity arrays, as it is designed to describe cases where the output of one quantum system serves as the input of the next. It has the further technical advantage that it makes two standard approximations at an early stage, so that the concomitant simplifications appear right from the start. One approximation is equivalent to the Wigner-Weisskopf approximation for spontaneous decay of an atom and yields a simple decay rate for each cavity field. The other approximation assumes an isolated discrete cavity mode with  a well-defined resonance frequency [cf. the two Eqs.~(\ref{inout}) and (\ref{kappa})].

In addition, we will use the quantum trajectory method \cite{gardiner2004quantum,zoller1987,carmichael1993open,dum1992monte,molmer1993monte,plenio1998quantum}, which is well suited to describe open quantum systems. Thus, both dissipation due to spontaneous emission of an atom or losses inside the cavity, and detection by photo detectors are easily incorporated. The quantum evolution of the open system consists of two parts in this picture:  there are discrete quantum jumps at random times (occurring with certain probabilities), corresponding to the detection of single photons or spontaneous emission events, and a jump-free evolution where no detection or emission event takes place. 
On average the combined evolution is identical to that obtained from a Master equation for the density matrix of the same system.

The quantum-optics literature and the classical optics literature use different descriptions of the coupled cavity systems.
In fact, the input-output formalism leads to different  equations than those used in the classical theory. (This difference is due to the approximations used in the input-output theory, rather than due to a difference between quantum and classical physics. The latter difference shows up in the photon statistics and could be revealed by measuring higher-order correlation functions, such as $g^{(2)}$ \cite{mandel} or time correlations \cite{aspect1980}). It is useful to state here the relation between the two descriptions. For this it suffices to consider the simplest case of one mode inside a lossless ring resonator, with a single input and a single output field (see FIG.~1).
\begin{figure}[h]
\includegraphics[width=2.5in,height=2.1in]{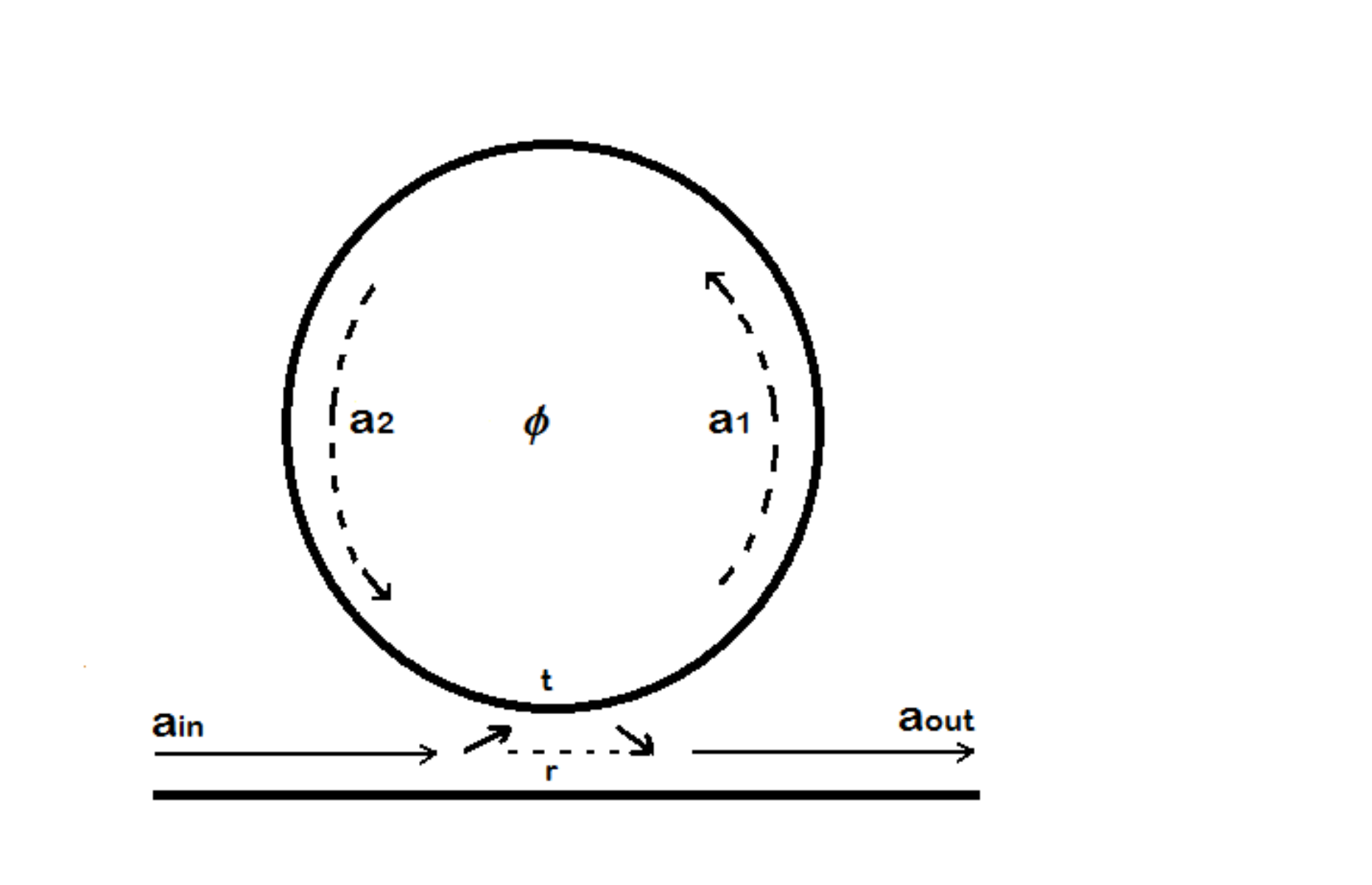}
\captionsetup{
  margin=0em,
  justification=raggedright,
  singlelinecheck=false
}\caption{Classical description of an empty lossless ring resonator coupled to an optical fiber. Analogous to a beam splitter configuration it is assumed that part of the input field transmits to the ring resonator and the rest reflects back into the fiber. This  process is described by transmission  and reflection coefficients $t$ and $r$, respectively, and leads to the relations $a_1=ta_2+ira_{{\rm in}}$, and $a_{{\rm out}}=ira_2+ta_{{\rm in}}$. Moreover, upon one roundtrip the field amplitude gains a phase $\phi$, that is, $a_2=e^{i\phi}a_1$.  Eliminating the cavity modes yields the relation between output and input field amplitudes as given in Eq.~(\ref{crow}), up to an irrelevant overall minus sign.
For the {\em quantum} description one writes down an approximate Hamiltonian that couples a {\em single} cavity mode (this is one approximation) with resonance frequency $\omega_c$ to a continuum of fiber modes with a coupling rate that is assumed constant over the relevant range of continuum frequencies around resonance (this is a second approximation). After elimination of the fiber modes, one obtains an approximation to Eq.~(\ref{crow}) valid quantum mechanically, viz.~(\ref{inout}).
}\label{Fig0}
 \end{figure}
The classical theory gives this relation between input and output at some fixed frequency $\omega$ in the steady state:
\be\label{crow}
a_{{\rm out}}=\frac{-t+e^{i\phi}}{1-te^{i\phi}}a_{{\rm in}},
\ee
in terms of a roundtrip phase difference $\phi$ and real reflection and transmission coefficients $r,t$ which satisfy $r^2+t^2=1$. Input-output theory gives, instead, in the same situation
\be\label{inout}
a_{{\rm out}}=\frac{\kappa/2+i(\omega-\omega_c)}{\kappa/2-i(\omega-\omega_c)}a_{{\rm in}},
\ee
with $\omega_c$ the  cavity resonance frequency, and with $\kappa$ the cavity decay rate for energy.
The expression (\ref{inout}) 
can, alternatively, be found from (\ref{crow}) in the regime where $r\ll 1$ and $\phi\ll 1$, by approximating $t=\sqrt{1-r^2}\approx1-r^2/2$, $e^{i\phi}\approx 1+i\phi$, and identifying
\be\label{kappa}
\kappa\tau=r^2; \,\,\,\phi=(\omega-\omega_c)a\tau,
\ee
where $\tau$ is the roundtrip time of a photon in the ring cavity and $a$ is the radius of the ring cavity. From $r\ll 1$ it follows that $\tau\ll 1/\kappa$, and we can translate this to the statement that input-output theory is valid only on time scales long compared to the cavity round-trip time. Similarly, the approximation $\phi\ll 1$ implies frequencies should not be too far from resonance, as compared to the inverse cavity round-trip time.

Our main aim is to use the final expression for $a_{{\rm out}}$  to display the trapping and delaying effects of the cavity array in a direct way: we will obtain the time-dependent probabilities of detecting photons at the output, as well as their time-dependent spectrum, i.e., the time-dependent probabilities of detecting photons after the output field has traversed a frequency filter \cite{eberly1977time}.

The rest of the paper is organized as follows. We first consider a single two-level atom, assumed to start off in the excited state, in a single ring resonator. This Section (II) is meant mainly to establish notation,
and to present the input-output formalism and the quantum trajectory method in a relatively simple case.

Even in this simple case, by introducing detectors with a finite bandwidth, we enter in principle the territory of cascaded quantum systems, since the output of the atom-cavity system serves as the input of the detector, part of which can itself be modelled as a cavity. This will allow us to calculate the time-dependent spectrum of the single photon produced by the initially excited atom in Sec.~II.C. 

Then, in the next Section, III, we consider the case of an empty cavity driven by the atom-cavity system.
The presence of this second cavity opens the possibility of trapping and delaying the photon produced by the atom inside the first cavity. More precisely, the frequency component resonant with the second cavity will be delayed. This will show up in the time-dependent spectrum of that photon.

The two-cavity case can be easily generalized to any number of empty cavities. We still are able to obtain analytical solutions for this case, and we display results for two, three, four and five cavities in Section IV.

In the final results section, Section V, we treat the atom more realistically, pertaining to a situation closer to what one would experimentally implement: a three-level $\Lambda$ system, with an additional laser driving a Raman transition between two (hyperfine) ground states through an off-resonant intermediate excited state. This allows one to deterministically produce a cavity photon, with some control over the lineshape produced, while avoiding spontaneous emission. We did not find analytical solutions for this case, but the equations allow for straightforward numerical solutions, which in turn yield time-dependent spectra, among other quantities of interest.

\section{Single Two Level Atom Coupled to A Lossy Ring Resonator}
\subsection{Model and Hamiltonian}
\begin{figure}[h]
\includegraphics[width=2.4in,height=2in]{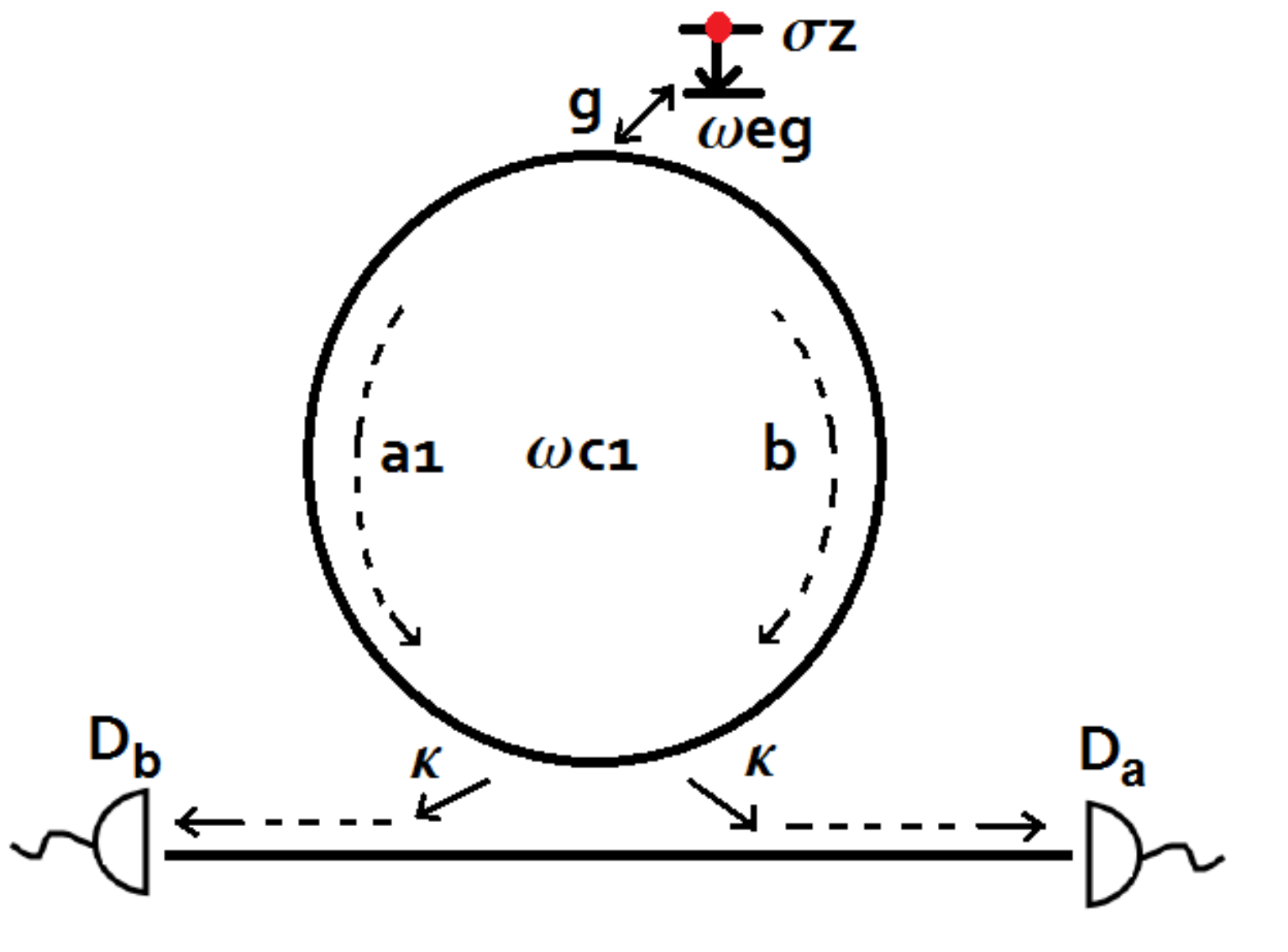}
\captionsetup{
  margin=0em,
  justification=raggedright,
  singlelinecheck=false
}
  \caption{An initially excited two-level atom can emit a photon into one of two counterpropagating modes of a lossy ring resonator. The photon leaks out at a rate $\kappa$ and is detected by one of two frequency-selective detectors with spectral width $\Gamma$.}\label{Fig1}
 \end{figure}
We start with the system depicted in FIG.~\ref{Fig1}. We have a ring resonator with decay rate $\kappa_{1}$ coupled to an initially excited two level atom (with resonance frequency $\omega_{eg}$) with complex coupling rate $g$. We neglect spontaneous emission  because we have in mind ultimately applying the formalism to a three-level atom in a configuration where spontaneous emission can indeed be ignored, see Section V.  Thus, when the atom de-excites it will excite one of two counter propagating modes in the resonator described by the annihilation operators $\hat{a}_{1}$ and $\hat{b}$, respectively. Mode $\hat{a}_{1}$ couples to the atom with the coupling rate $g$ and $\hat{b}$ couples with the coupling rate $g^{\ast}$, where the phase of $g$ describes the atomic location on the circumference of the resonator as in \cite{dayan2008photon} \footnote{This is a recent experiment  in the single-photon regime using a single atom inside a single cavity, where the nonlinearity introduced by the presence of the atom is exploited to change an incoming laser beam of light (with {\em Poissonian} photon-number statistics) into reflected light that is {\em antibunched}, characteristic of single photons}. Light from the ring cavity couples to an optical fiber, which is modelled to have a continuum of modes. 
The cavity interaction with the continuum of modes in the fiber can be incorporated simply through input field operators $\hat{a}_{{\rm in}}$ and $\hat{b}_{{\rm in}}$ (as introduced in \cite{gardiner2004quantum}). We take  the system Hamiltonian to have the form
\begin{equation}\label{HS}
\begin{split}
& \hat{H}_s=-\hbar\omega_{eg}\hat{\sigma}_{-}\hat{\sigma}_{+}+\hbar\omega_{c1}(\hat{a}_{1}^{\dagger}\hat{a}_{1}+ \hat{b}^{\dagger}\hat{b}) +\hbar(g\hat{a}_{1}^{\dagger}\hat{\sigma}_{-}\\ & +g^{\ast}\hat{a}_{1}\hat{\sigma}_{+})+\hbar(g^{\ast}\hat{b}^{\dagger}\hat{\sigma}_{-} +g\hat{b}\hat{\sigma}_{+})
-i\hbar\sqrt{\kappa}_{1}(\hat{a}_{1}^{\dagger}\hat{a}_{{\rm in}}\\
&-\hat{a}_{{\rm in}}^{\dagger}\hat{a}_{1}+\hat{b}^{\dagger}\hat{b}_{{\rm in}}-\hat{b}_{{\rm in}}^{\dagger}\hat{b}).
\end{split}
\end{equation}
We have assumed here the usual rotating wave approximation. In addition, for simplicity we assumed there is no direct coupling between the two intra-cavity modes (there is the indirect coupling through the atom). We also assumed a single resonance frequency $\omega_{c1}$ for both cavity modes. 
Non vanishing commutation relations are:
 $[\hat{\sigma_{+}},\hat{\sigma_{-}}]=\hat{\sigma_{z}}$, $[\hat {a}_{1},\hat {a}_{1}^{\dagger}]=1$ and  $[\hat {b},\hat {b}^{\dagger}]=1$. 
 The input field operators are not dimensionless and satisfy different commutation relations, namely
$[\hat{a}_{{\rm in}}(t),\hat{a}_{{\rm in}}^{\dagger}(t')]=\delta(t-t')$ and the analogous relation for $\hat{b}_{{\rm in}}$. 
 Corresponding to the input operators there are two output field operators then, denoted by $\hat{a}_{{\rm out}}$ and $\hat{b}_{{\rm out}}$ which are related to the input fields and 
 intra-cavity modes through the input-output relations (obtained by formally solving the Heisenberg equations for the fiber modes) as:
 \begin{subequations}
  \label{about}
 \begin{eqnarray}
 \hat{a}_{{\rm out}}&=&\hat{a}_{{\rm in}}+\sqrt{\kappa_{1}}\hat{a}_{1},\\
 \hat{b}_{{\rm out}}&=&\hat{b}_{{\rm in}}+\sqrt{\kappa_{1}}\hat{b}.
 \end{eqnarray}
 \end{subequations}
 These output fields physically correspond to the electric fields at the point where the fiber couples to the two cavity modes.
If we denote by $\ket{\Psi}$ the initial state of the global system (atom, cavity and fiber) we have $\hat{a}_{{\rm in}}\ket{\Psi}=0$ and $\hat{b}_{{\rm in}}\ket{\Psi}=0$, as initially there is no photon in the fiber. For this reason we will sometimes suppress terms containing the input fields
in normally ordered expressions, since those terms do not contribute to expectation values.

\subsection{Quantum trajectory analysis}
Coupling of the cavity to the output fields in the fiber makes the system open as soon as we have eliminated the continuum modes via Eqs.~(\ref{about}). In order to describe such a dissipative system we apply the quantum trajectory (or quantum jump) method \cite{gardiner2004quantum,zoller1987,carmichael1993open,dum1992monte,molmer1993monte,plenio1998quantum}.
\subsubsection{Occurrence of a Jump }
 In this method we can think of two (fictitious or actual) detectors $D_{a}$ and $D_{b}$, one placed at the right end and the other at the left end of the fiber as shown in Fig.[\ref{Fig1}]. $D_{a}$ detects $\hat{a}_{{\rm out}}$ and $D_{b}$ detects $\hat{b}_{{\rm out}}$. We assume the detectors to be perfectly efficient so that if a photon leaks out of the cavity one of the detectors will detect it by making a click.

In a given small time interval around a time $t$,  there are two possibilities then, either one of the detectors clicks or none does. We consider the case of a detector clicking first, indicating that a jump has occurred. In our case we have two detectors and hence also  two jump operators. We denote these by $\hat{J}_{a}$ and $\hat{J}_{b}$, and we have $\hat{J}_{a}=\hat{a}_{{\rm out}}$ and $\hat{J}_{b}=\hat{b}_{{\rm out}}$. 
The state of the system of cavity modes and atom before the jump, , $\ket{\psi}$, is reset after a jump of type $j=a,b$ by the transformation
\begin{equation}
\ket{\psi}\mapsto
\frac{\hat{J}_j\ket{\psi}}{\sqrt{\Pi_j}}.
\end{equation}
The normalization factor $\Pi_j$ appearing here is in fact a probability density per time, defined as follows:
 The probability for a jump occurring during the infinitesimal time interval $[t,t+dt]$ is given by:
\begin{equation} 
P_{j}(t)={\bra{\psi}}\hat{J_{j}}^{\dagger}
\hat{J_{j}}{\ket{\psi}}dt=:\Pi_j dt,
\end{equation}
for $j=a,b$.

 \subsubsection{System dynamics if no jump occurs and the density operator}
 According to the Quantum Jump Theory, when no detector click occurs, the system dynamics follows a non-unitary evolution. The non-Hermitian ``Hamiltonian'' $\hat{H}_{NH}$ that drives this evolution can be written as the sum of the standard (Hermitian) Hamiltonian of (\ref{HS}) and terms describing decay constructed from the jump operators,
\begin{equation}\label{AHH}
\hat{H}_{NH}=\hat{H}_{s}-i\sum_{j=a,b}\hat{J}_j^\dagger\hat{J}_j/2.
\end{equation}
The system dynamics during the time of no jump is governed by the following non-unitary Schr\"odinger equation:
\begin{equation}\label{NUSE}
i\hbar\frac{d\ket{\tilde{\psi}(t)}}{dt}=\hat{H}_{NH}{\ket{\tilde{\psi}(t)}}.
\end{equation}
Here, $\ket{\tilde{\psi}(t)}$  is a pure state, but it is not normalized (its norm decays in time). In our case it can be written as a linear combination of the different possibilities of finding the photon in the system before being detected as:
\begin{equation}\label{NJS}
\ket{\tilde{\psi}(t)}=c_{e}(t)\ket{e,0,0}+c_{1}(t)\ket{g,1,0}+c_{2}(t)\ket{g,0,1},
\end{equation}
where we are using the notational conventions that the first slot in the ket denotes the state of the atom, and the second and third slots give the number of photons in the intra-cavity modes, $a_{1}$ and $b$, respectively. 
Note there is no term corresponding to a photon inside the fiber (the fiber modes were effectively eliminated when we solved the Heisenberg equations for the fiber-mode operators, see Eq.~(\ref{about})).

Using  $\eqref{AHH}$ and $\eqref{NJS}$  in $\eqref{NUSE}$ we get three coupled differential equation describing the time evolution of the probability amplitudes which can be easily solved using the Laplace transform. In Laplace space the equations for the amplitudes are algebraic and appear as
\begin{subequations}
\begin{eqnarray}
sC_{e}(s)+igC_{1}(s)+ig^{\ast}C_{2}(s)&=&1,\\
\Bigg( s+i\Delta+\frac{\kappa_{1}}{2} \Bigg)C_{1}(s)+ig^{\ast}C_{e}(s)&=&0,\\
\Bigg( s+i\Delta+\frac{\kappa_{1}}{2} \Bigg)C_{2}(s)+igC_{e}(s)&=&0,
\end{eqnarray}
\end{subequations}
with $\Delta=\omega_{c1}-\omega_{eg}$. $C_{j}(s)$ is the Laplace transform of $c_{j}(t)$ with $j=1,2,e$ and we have used the initial conditions $c_{e}(t=0)=1$ and $c_{i}(t=0)=0$ for $ i=1,2$. Solving these equations and then taking the inverse Laplace transform we arrive at the following analytic results for the amplitudes
\begin{subequations}
\begin{eqnarray}
c_{e}(t)&=&\frac{2e^{-(\frac{\kappa_{1}}{4}+i\frac{\Delta}{2})t}}{\alpha}\Bigg((i\Delta+\kappa_{1})\sinh(\alpha t)+\alpha\cosh(\alpha t)\Bigg),\nonumber\\
\label{ce}\\
\label{c1}
c_{1}(t)&=&\frac{-2ig^{\ast}}{\alpha}\Bigg(2e^{-(\frac{\kappa_{1}}{4}+i\frac{\Delta}{2})t}\sinh(\alpha t)\Bigg),\\
\label{c2}
c_{2}(t)&=&\frac{-2ig}{\alpha}\Bigg(2e^{-(\frac{\kappa_{1}}{4}+i\frac{\Delta}{2})t}\sinh(\alpha t)\Bigg),
\end{eqnarray}
\end{subequations}
where 
\[\alpha=\frac{\sqrt{\kappa_{1}^{2}+4i\kappa_{1}\Delta-4\Delta^{2}-32g^{2}}}{4}.\]
Note that $\alpha$ is a complex number so the amplitudes calculated above in Eqs.~(\ref{ce})--(\ref{c2}) are not purely decaying functions but may show oscillations, as we will confirm explicitly later.

\begin{figure*}
\begin{center}
\begin{tabular}{cccc}
\subfloat{\includegraphics[width=7cm,height=5cm]{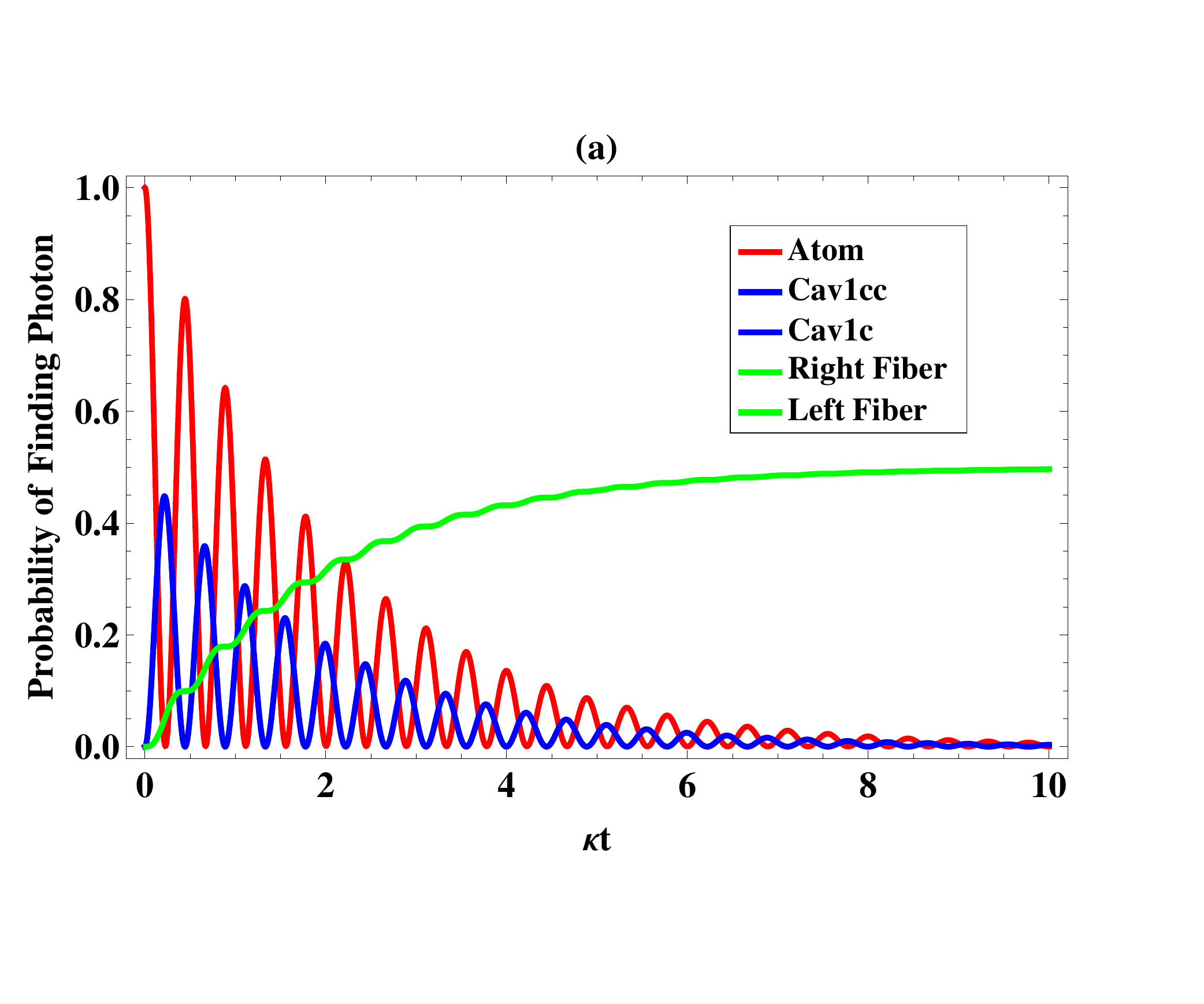}} & 
\subfloat{\includegraphics[width=7cm,height=5cm]{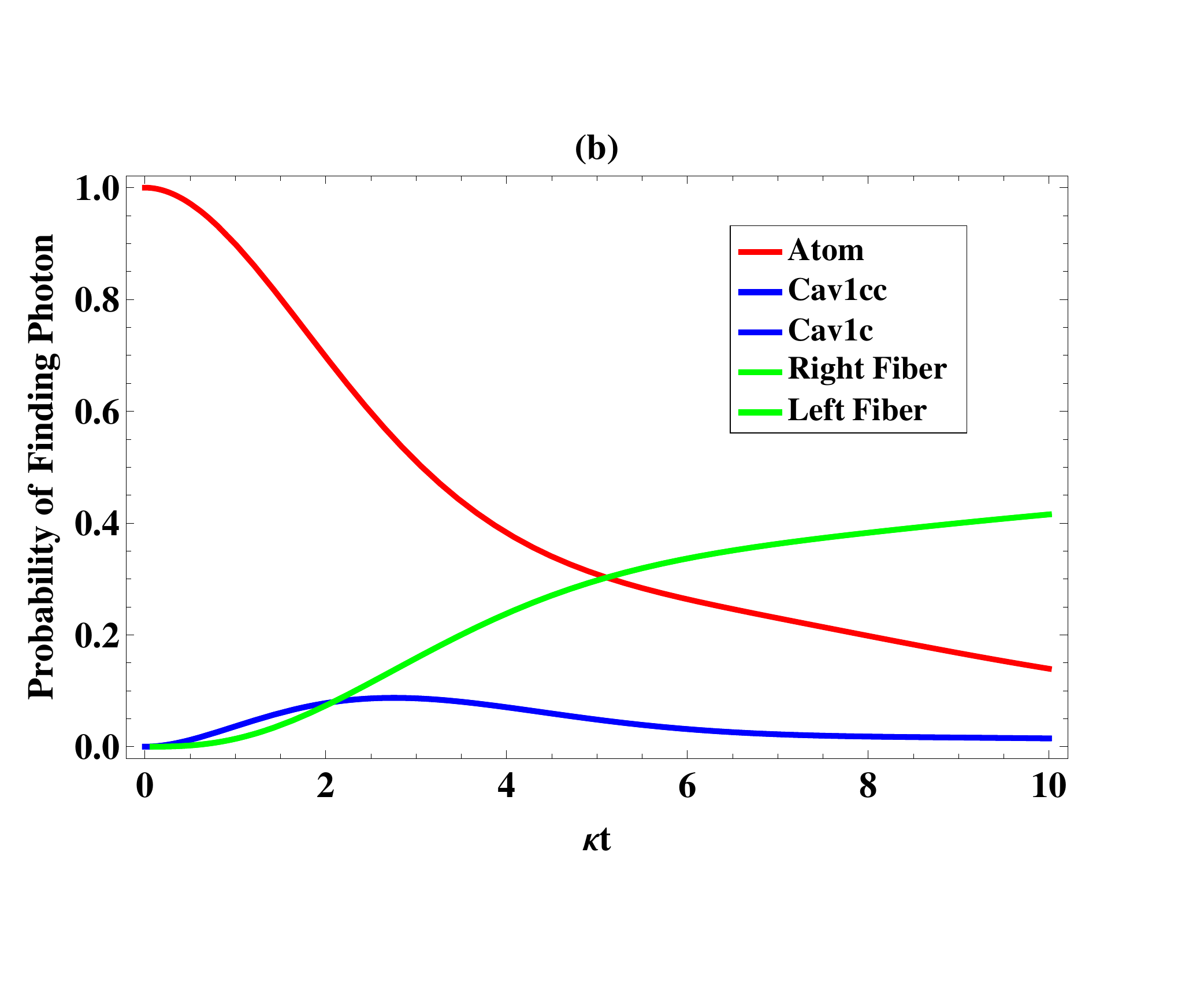}}\\
\end{tabular}
\captionsetup{
  margin=1em,
  justification=raggedright,
  singlelinecheck=false
}
\caption{Time evolution of probabilities of finding the single excitation in the atom, the cavity counter clockwise (1cc) mode, the cavity clockwise (1c) mode, and the right fiber and left fiber continuum modes. (a) Strong coupling regime: $|g|/\kappa_{1}=5, \Delta_{c1}/\kappa_{1}=(\omega_{c1}-\omega_{eg})/\kappa_{1}=0.5$. The oscillatory behaviour in the plots is the manifestation of the single-photon Rabi oscillation. (b) Weak coupling regime $|g|/\kappa{1}=0.25,   \Delta_{c1}/\kappa_{1}=(\omega_{c1}-\omega_{eg})/\kappa_{1}=0.5$. Here we see the non-oscillatory, monotonically decaying behaviour for the atomic probability. Note that after sufficiently long times, $\kappa_{1} t\simeq 10$, the probability of finding the photon in the left and right fiber modes approaches $0.5$, while all other probabilities die out.}\label{Fig2}
\end{center}
\end{figure*}

In our case, since we start with just one excitation in the system, and no external driving, there can be at most a single quantum jump.
After recording that jump the system's previous (unnormalized) state $\tilde{\ket{\psi}}$ will collapse to $
\hat{J}_{a,b}\tilde{\ket{\psi}}\rightarrow\ket{g,0,0}$.
In our special case, the state after the jump is independent of what the old state was, and independent of which of the two possible types of jump occurred. 
Following the quantum trajectory method we can, therefore, construct the total density operator $\hat{\rho}(t)$ describing the state of the system for all times, by performing an ensemble average over the two different types density operators, one indicating that a jump has occurred and the other with no jump, i.e., 
\begin{equation}
\hat{\rho}(t)=\ket{\tilde{\psi}(t)}\bra{\tilde{\psi}(t)}+P(t)\ket{g,0,0}\bra{g,0,0}.
\end{equation}
Here $P(t)=P_a(t)+P_b(t)$ is the probability of the occurrence of a jump (of either type) at time t. 
From the density operator defined above we can work out the time evolution of the probabilities of finding the initial excitation in the atom ($P_{e}(t)$), in the cavity modes ($P_{1}(t),P_{2}(t)$) and in the left and right fiber continua ($P_{k1}(t),P_{k2}(t)$),
\begin{subequations}
\begin{eqnarray}
P_{e}(t)&=&\Tr[\hat{\rho}(t)\ket{e,0,0}\bra{e,0,0}]=|c_{e}(t)|^2, \\
P_{1}(t)&=&\Tr[\hat{\rho}(t)\ket{g,1,0}\bra{g,1,0}]=|c_{1}(t)|^2,\\
P_{2}(t)&=&\Tr[\hat{\rho}(t)\ket{g,0,1}\bra{g,0,1}]=|c_{2}(t)|^2,\\
 P_{k1}(t)&=&\kappa_{1}\int_0^t\Tr[\rho(t)\hat{a}_{1}^{\dagger}\hat{a}_{1}]dt=\kappa_{1}\int_0^t|c_{1}(t')|^2dt',\nonumber\\
 \\
P_{k2}(t)t&=&\kappa_{1}\int_0^t\Tr[\rho(t)\hat{b}^{\dagger}\hat{b}]dt=\kappa_{1}\int_0^t|c_{2}(t')|^2dt'.\nonumber\\
\end{eqnarray}
\end{subequations}
In Fig.~\ref{Fig2} we have plotted these probabilities in the strong $(g>\kappa_{1})$ and weak $(g<\kappa_{1})$ coupling regimes.

 In the strong coupling regime our results agree with the Wigner-Weisskopf approach  discussed in \cite{cui2006emission}, where the various probabilities display the well-known single-photon Rabi oscillations, indicating the (almost) reversible energy excitation exchange between emitter and cavity.  The amplitude of the oscillations decays  due to the lossy nature of the cavity. 
 
 In the weak coupling case  we arrive at the usual  irreversible decaying behaviour as found out in \cite{auffeves2008spontaneous} where the cavity behaves as the relaxation channel for the excited atom.

\subsection{Emission spectrum}
\begin{figure*}[h]
\begin{center}
\begin{tabular}{cccc}
\subfloat{\includegraphics[width=6cm,height=5cm]{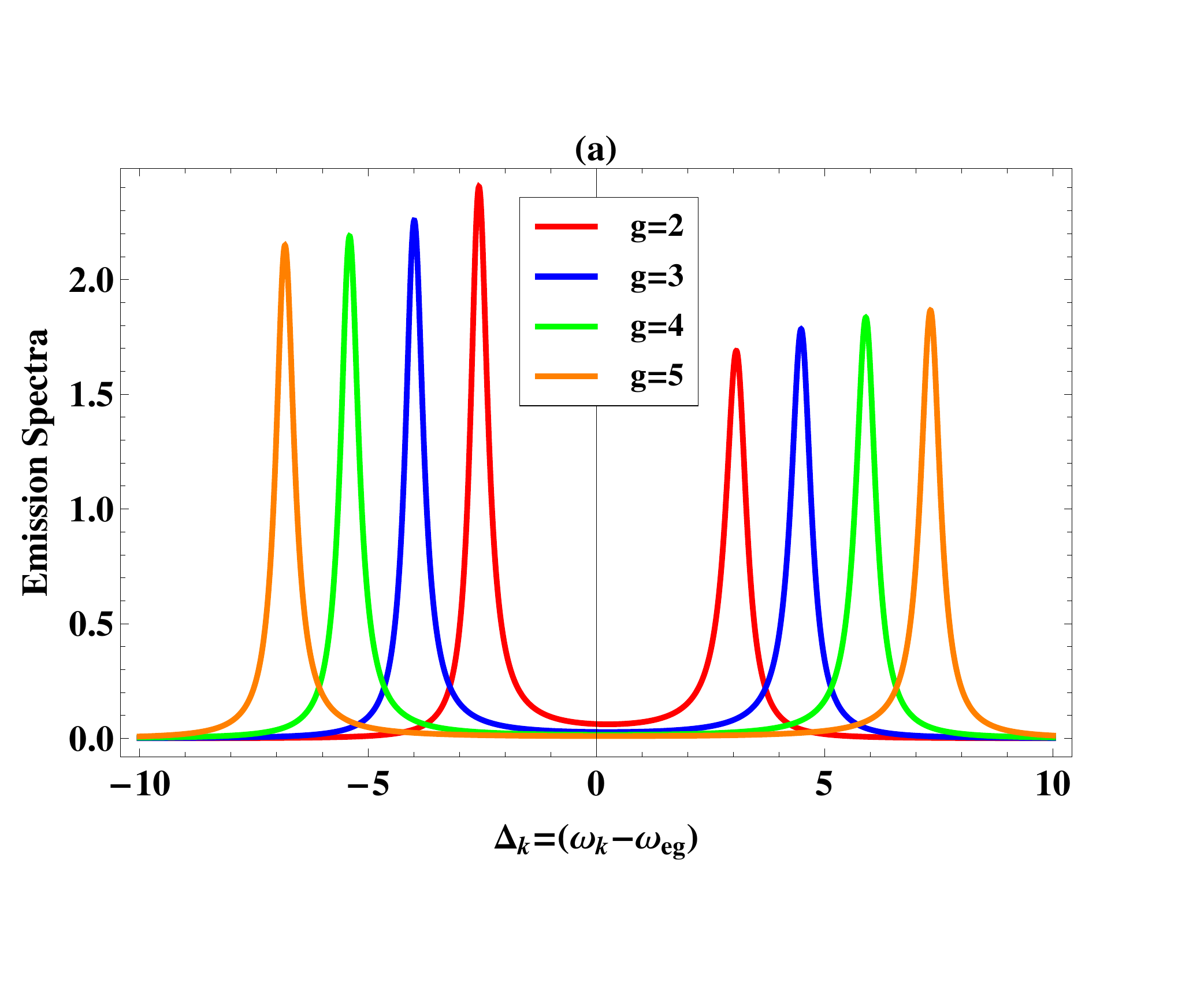}} & 
\subfloat{\includegraphics[width=6cm,height=5cm]{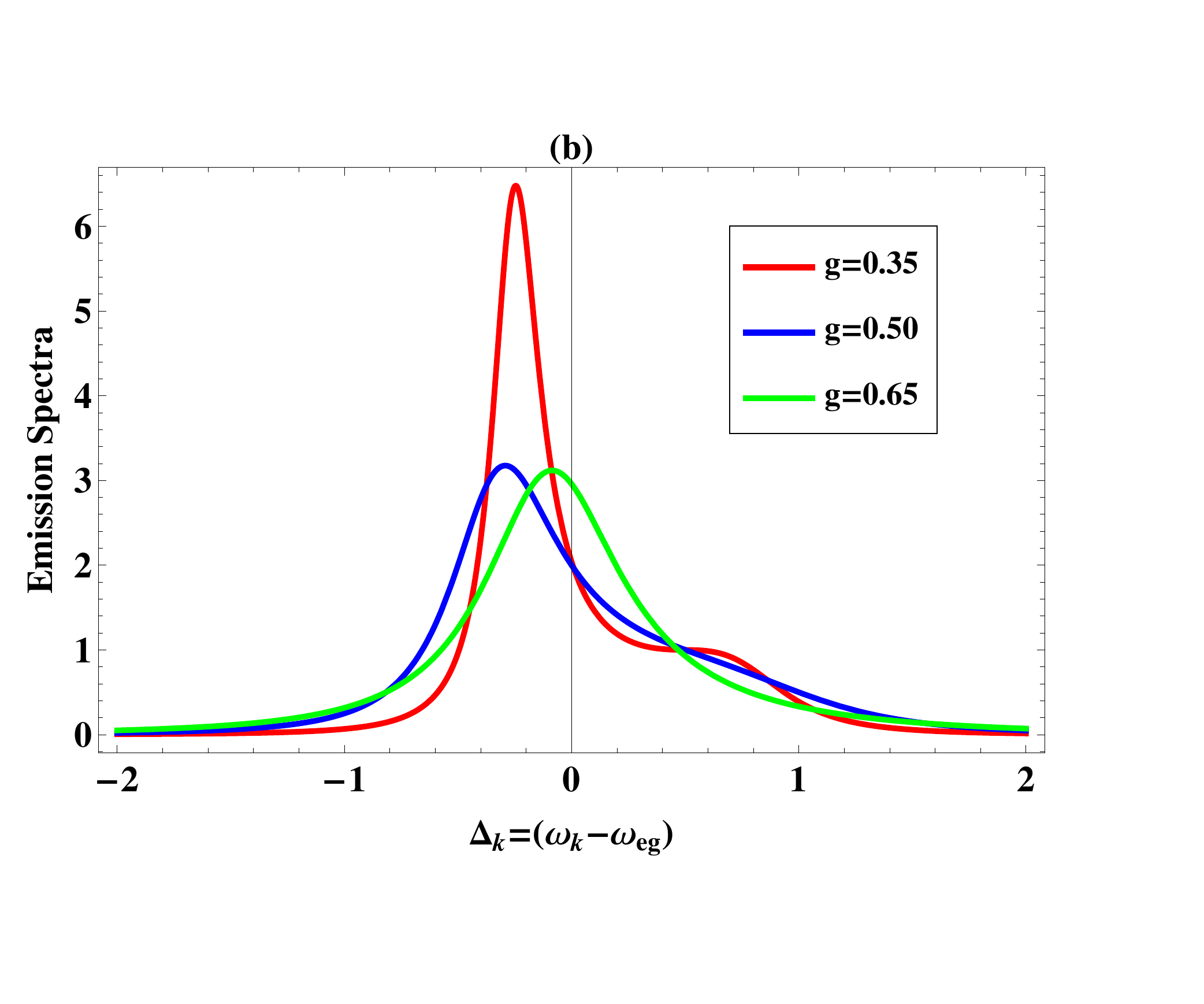}}
\end{tabular}
\captionsetup{
  margin=1em,
  justification=raggedright,
  singlelinecheck=false
}
\caption{Emission spectra (as functions of $\Delta_k$ in units of $\kappa_1$) recorded by a detector with bandwidth $\Gamma/\kappa_{1}=0.25$. (a) Strong coupling regime,  $\Delta/\kappa_{1}=0.5$ and varying values of $|g|/\kappa_{1}$. Note that the single-photon Rabi splitting equals $2\sqrt{2}|g|$. The asymmetry in the heights of the peaks is due to a nonzero detuning between atom and cavity (which breaks the symmetry under $\omega_k\leftrightarrow -\omega_k$). (b) Emission spectra with varying values of $|g|/\kappa_{1}$ remaining in the weak coupling regime and $\Delta/\kappa_{1}=0.5$.  }\label{Fig3}
\end{center}
\end{figure*}

The quantum trajectory method for cascaded quantum systems can be used to obtain the (time-dependent) light spectrum emitted by the system of interest \cite{tian1992quantum,havukainen1998open}. Here we calculate this spectrum with the help of a method discussed in great detail in \cite{carmichael2007statistical}. We are going to think about taking measurements on the output fields through first coupling the output light into a  frequency filter, and only then counting photons. The atom-cavity system and the frequency filter comprise the cascaded system as depicted in FIG.~\ref{Fig1}. 
This approach \cite{carmichael2007statistical}  leads us, in fact, to the standard results for the classical optical spectrum, introduced by Eberly and Wodkiewicz \cite{eberly1977time}, except that the classical field amplitudes are replaced by quantum annihilation and creation operators. We arrive at the following expression for the spectrum emitted by our atom-cavity system:
\begin{equation}\label{TDPS}
\begin{split}
&N(t;\Delta_{k},\Gamma)=\\
&\Gamma\int_{0} ^t\int_{0} ^t e^{-(\Gamma-i\Delta_{k})(t-t_{1})}e^{-(\Gamma+i\Delta_{k})(t-t_{2})}\times\\
&\langle\hat{a}^{\dagger}_{out}(t_{1})\hat{a}_{out}(t_{2})\rangle dt_{1}dt_{2}.
\end{split}
\end{equation}
 As before, the output $\hat{a}_{out}$ from the atom-cavity system is given by $\sqrt{\kappa_{1}}\hat{a}_{1}$. This expression (\ref{TDPS}) defines the spectrum, with $N$ being the counting rate of the detector after having frequency filtered the light with a filter with bandwidth $\Gamma$ and detuning $\Delta_{k}=\omega_k-\omega_{eg}$ from the atomic transition frequency. In order to connect to previous work (which used different methods, see below), we can also integrate Eq.~(\ref{TDPS}) over  time  to obtain the ``synthesized'' spectrum,
 \begin{equation}\label{SPS}
 N_S(t;\Delta_k,\Gamma)=
 \int_0^t N(t';\Delta_k,\Gamma)dt'.
 \end{equation}
 In the limit of $t\rightarrow\infty$ this quantity would equal the spectrum for a stationary process as obtained from the Wiener-Khinchine theorem,
   $P_{D_a}({\Delta_{k}})$, which turns out to be
 \begin{equation}
 P_{D_a}({\Delta_{k}})=
\frac{4|g|^2\kappa_{1}\Gamma}{[4|g|^2-2\Delta_{k}(\Delta_{k}+\Delta)]^2+\kappa_{1}^2\Delta_{k}^2}.
\end{equation}
(In the next Sections, covering multiple coupled cavities, we will present time-dependent spectra, as defined by (\ref{TDPS}), as well as its time-integrated version.)
In Fig.~\ref{Fig3} we have plotted the synthesized spectrum emitted by the lossy cavity. It is a doublet having two resonances one around $\omega=\omega_{eg}$ and the other at $\omega=\omega_{c1}$, both shifted from the bare resonance by an amount depending on the value of $|g|^2$.
 This separation in frequency space is the one-photon Rabi splitting whose value is given by $2\sqrt{2} |g|$. Our results are consistent with those of Refs.~\cite{shen2009theory, auffeves2008spontaneous,carmichael1989subnatural} where the emitted spectrum was calculated using different methods, viz., real-space quantization \cite{shen2009theory}, the Wiener-Khinchine theorem \cite{auffeves2008spontaneous}, and the quantum regression theorem \cite{carmichael1989subnatural}.

\section{Empty ring resonator driven by an atom-cavity system}
\begin{figure}[h]
\includegraphics[width=3.25in,height=2in]{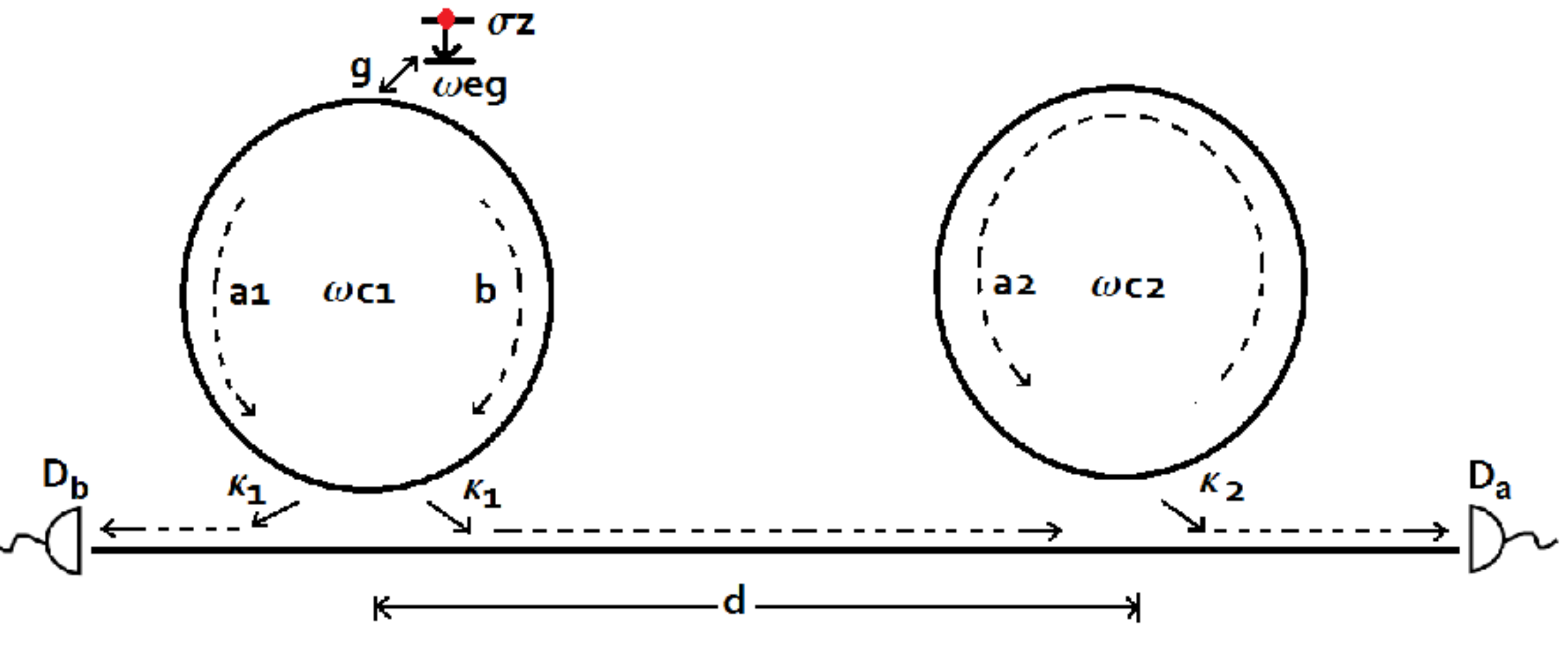}
\captionsetup{
  margin=0em,
  justification=raggedright,
  singlelinecheck=false
}  \caption{An empty ring resonator driven by an atom-cavity system. The input of the second cavity equals the output field of the first cavity, delayed by a time $\tau_d$.}\label{Fig4}
 \end{figure}

We consider now the system displayed in Fig.~[\ref{Fig4}].
The system of the previous Section, a single two level atom coupled to a lossy ring cavity (with parameters as before) is coupled to a second ring cavity that is empty. The latter may have a different resonance frequency $\omega_{c2}$ than the first cavity, and a different decay rate $\kappa_{2}$, as shown in Fig.~\ref{Fig4}. The cavities are separated by a distance $d$ which causes a time delay $\tau_{d}=d/c$ for the light to propagate from one cavity to the other, $c$ here being the group velocity of the light in the fiber, assumed constant around both cavity resonance frequencies. This delay can be eliminated simply by defining ``time delayed" operators. For instance, we define $\hat{a}_{2}(t):=\hat{a}_{2}(t-\tau_{d})$ etcetera \cite{cirac1997quantum}. We can do this so simply because we assume the first cavity is not driven by fields  from the second cavity.

Following the quantum trajectory approach for cascaded systems, the crucial ingredient is that the output of the first cavity is serving as the input of the second cavity. That is,
we have 
\be
\hat{a}_{{\rm in}}^{(2)}(t)=\hat{a}_{{\rm out}}^{(1)}(t),
\ee
where the same time arguments appear on the left and right-hand sides thanks to the elimination of the time delay $\tau_d$.
\subsection{Jump operators}
We have again two jump operators, $\hat{J}_a$ and $\hat{J}_b$, describing quantum jumps corresponding to clicks in detectors $D_a$ and $D_b$, respectively. Like before, we have $\hat{J}_b=\hat{b}_{{\rm out}}=\sqrt{\kappa_{1}}\hat{b}$, but $\hat{J}_a=\hat{a}_{{\rm out}}^{(2)}$ is now the output field from the {\em second} resonator. From input-output theory this output field has the form
\begin{equation}
\hat{a}_{{\rm out}}^{(2)}=\hat{a}_{{\rm in}}^{(2)}+\sqrt{\kappa_{2}}\hat{a}_{2}.
\end{equation}
Substituting $\hat{a}_{{\rm in}}^{(2)}=\hat{a}_{{\rm out}}^{(1)}
=\sqrt{\kappa_{1}}\hat{a}_{1}+\hat{a}_{{\rm in}}^{(1)}$ yields
\begin{equation}
\hat{J}_a=\sqrt{\kappa_{1}}\hat{a}_{1}+\sqrt{\kappa_{2}}\hat{a}_{2}+\hat{a}_{{\rm in}}^{(1)}.
\end{equation}
Now our jump operator $\hat{J}_a$ has three parts, reflecting the fact that the detector $D_a$ cannot distinguish  photons emitted by cavity 1 or by 2 or by the input field (in our case, the latter type of photons is absent, of course). 

The evolution of the system's state due to quantum jumps is essentially the same as before, as a jump can only take the system  to its ground state, with the atom in state $\ket{g}$ and no photons. We thus focus now on the jump-free evolution.
\subsection{Hamiltonian}
With this, the Hamiltonian of the system can then be divided into two parts. The first part is Hermitian and given by
\begin{equation}\label{HH}
\hat{H}_{H}=\hat{H}_{A}+\hat{H}_{B}-i\hbar\frac{\sqrt{\kappa_{1}\kappa_{2}}}{2}(\hat{a}_{2}^{\dagger}\hat{a}_{1}-
\hat{a}_{2}\hat{a}_{1}^{\dagger}),
\end{equation}
where the last term arises from 
$\hat{a}_{{\rm in}}^{(2)}$ driving the second cavity, and
$\hat{H}_{A}$  is given by Eq.~(\ref{HS}), and
\begin{equation}
\hat{H}_{B}=\hbar\omega_{c2}\hat{a}_{2}^\dagger\hat{a}_{2}.
\end{equation}
The second part of the non-Hermitian Hamiltonian is anti-Hermitian and is given, as before, by $-i\sum_{j=a,b}\hat{J}_j^\dagger\hat{J}_j/2$, which here  amounts to
\begin{equation}\label{AH}
\begin{split}
\hat{H}_{AH}&=-\frac{i\hbar}{2}(\kappa_{1}\hat{a}_{1}^{\dagger}\hat{a}_{1}
+\kappa_{1}\hat{b}^{\dagger}\hat{b}
+\kappa_{2}\hat{a}_{2}^{\dagger}\hat{a}_{2})\\
&-i\hbar\frac{\sqrt{\kappa_{1}\kappa_{2}}}{2}(\hat{a}_{2}^{\dagger}\hat{a}_{1}+
\hat{a}_{2}\hat{a}_{1}^{\dagger}).
\end{split}
\end{equation}
We note that some of the terms in Eqs.~(\ref{HH}) and (\ref{AH}) cancel out (in particular, the counter-intuitive term describing the process in which a photon is created in cavity 1 upon destruction of a photon in cavity 2 cancels) and what remains is a non-Hermitian Hamiltonian given by:
 \begin{equation}
\begin{split}
& \hat{H}_{NH}= \hat{H}_{A}+\hat{H}_{B}-\frac{i\hbar}{2}(\kappa_{1}\hat{a}_{1}^{\dagger}\hat{a}_{1}
+\kappa_{1}\hat{b}^{\dagger}\hat{b}\\
&+\kappa_{2}\hat{a}_{2}^{\dagger}\hat{a}_{2})
-i\hbar\sqrt{\kappa_{1}\kappa_{2}}\hat{a}_{2}^\dagger\hat{a}_{1}.
\end{split}
\end{equation}
The above Hamiltonian describes a unidirectional coupling between source and target such that a photon can be created in the second cavity by annihilating a photon in the first cavity but not the other way round. 

Including the second cavity in our system of interest increases the dimension of our truncated Hilbert space by one. During any time interval where no photon is detected, the unnormalized state of the discrete systems now be written as
\begin{equation}
\begin{split}
&\ket{\tilde{\psi}(t)}=c_{e}(t)\ket{e,0,0,0}+c_{1}(t)\ket{g,1,0,0}\\
&+c_{2}(t)\ket{g,0,1,0}+c_{3}(t)\ket{g,0,0,1}.
\end{split}
\end{equation}
As we did for operators, in the probability amplitudes we have absorbed the time delay, so that for instance $c_{2}(t):=c_{2}(t-\tau_{d})$ and so forth. These amplitudes can be worked out by the same procedure as discussed before. Analytic expressions for $c_{e}(t)$, $c_{1}(t)$ and $c_{2}(t)$ are exactly the same as in Eqs.~(\ref{ce}), (\ref{c1}) and (\ref{c2}) respectively, and $c_{3}(t)$ (where we display only the simpler case of $\omega_{c1}=\omega_{c2}\equiv\omega_{c}$,   $\kappa_{1}=\kappa_{2}\equiv\kappa$) is  given by:
\begin{equation}
\begin{split}
&c_{3}(t)=\frac{ig\kappa}{\sqrt{\alpha}(8|g|^2+2i\kappa\Delta)}\Bigg[4\sqrt{\alpha}e^{-(i\Delta+\frac{\kappa}{2})t} \\
& -4e^{-(\kappa/4+i\Delta/2)t}\lbrace{(-2i\Delta-\kappa)\sinh(\alpha t)}\\
&-4e^{-(\kappa/4+i\Delta/2)t}{\alpha\cosh(\alpha t)\rbrace}\Bigg],
\end{split}
\end{equation}
where $\alpha=\frac{\sqrt{\kappa+4i\kappa\Delta-4\Delta^{2}-32g^{2}}}{4}$ and $\Delta=\omega_{c}-\omega_{eg}$.

\begin{figure*}[h]
\begin{tabular}{cccc}
\subfloat{\includegraphics[width=6cm,height=5cm]{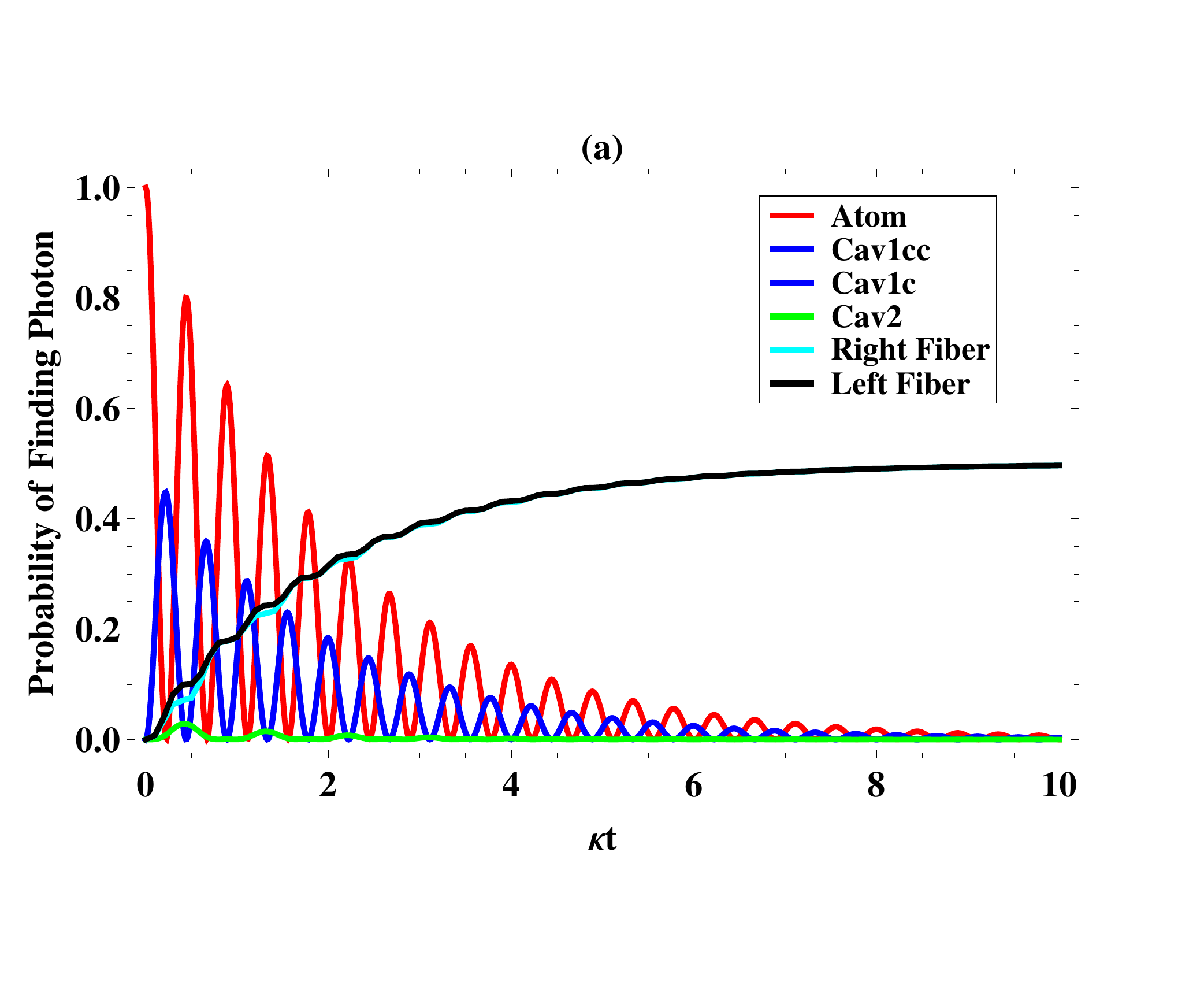}} & 
\subfloat{\includegraphics[width=6cm,height=5cm]{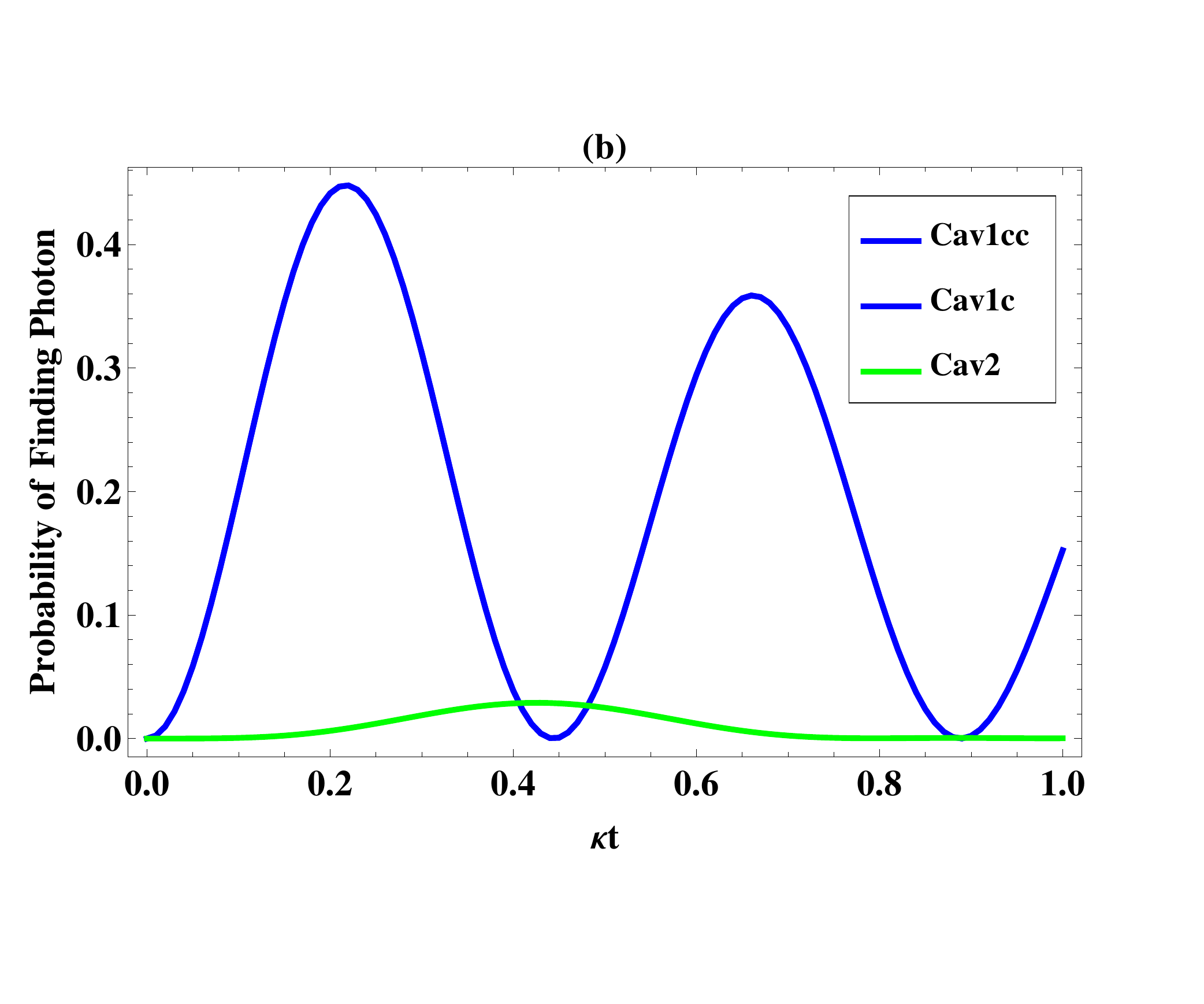}} &
\subfloat{\includegraphics[width=6cm,height=5cm]{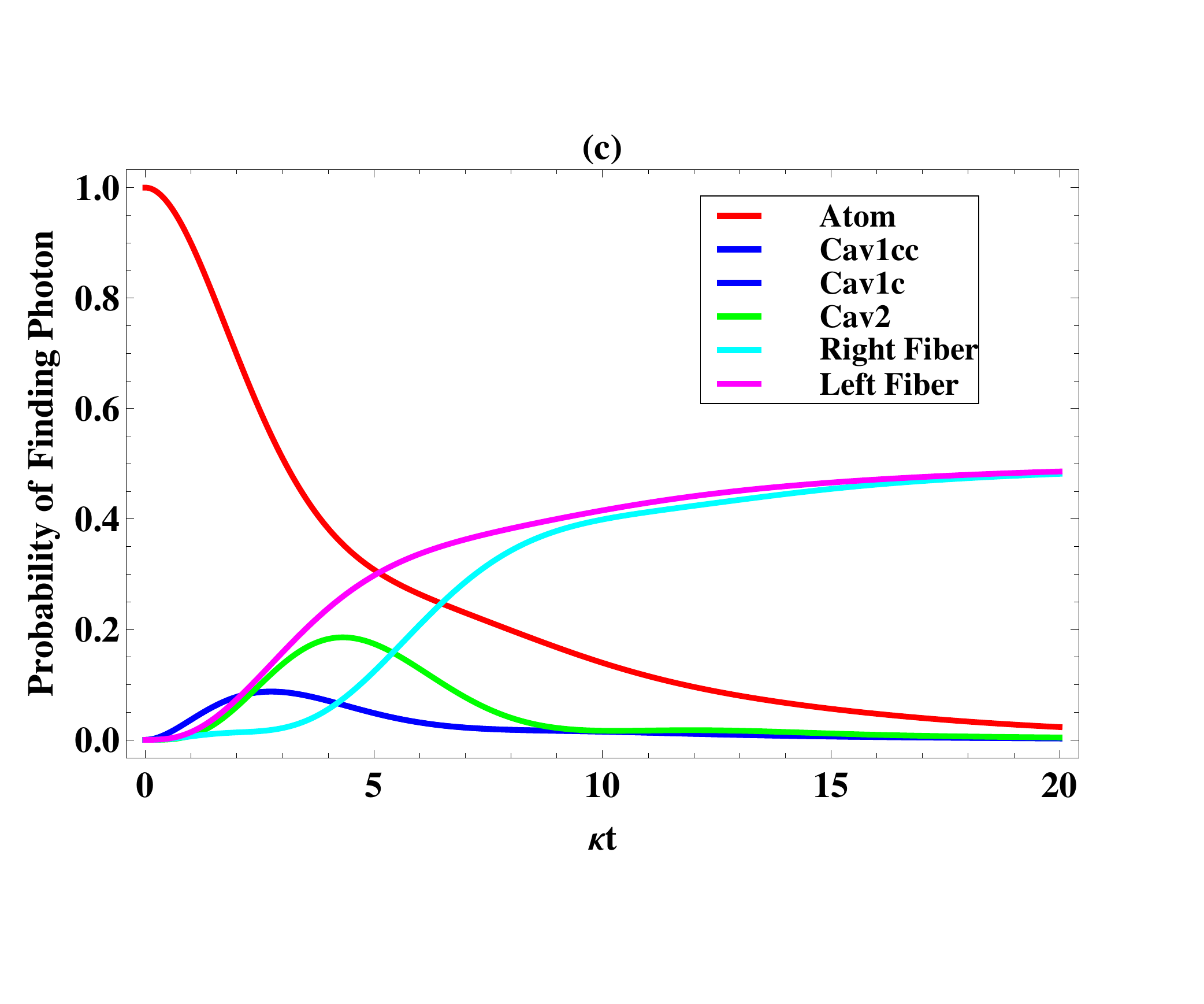}}\\
\end{tabular}
\captionsetup{
  margin=0em,
  justification=raggedright,
  singlelinecheck=false
}\caption{Probabilities of finding the excitation in the atom, in the 1st cavity clockwise-anticlockwise modes, in the 2nd cavity (Cav2) and in the left-right fibers. (a) Strong coupling regime, with $|g|/\kappa_{A}=|g|/\kappa_{B}=5, \Delta_{1}/\kappa_{A}=\Delta_{2}/\kappa_{B}=0.5$. Note here that the populations in left and right fiber modes are identical, so only one curve is visible. Fig.~(b) is an enlargement of Fig.~(a). It shows the photon time delay (of about $\sim$0.24$\kappa^{-1}$) between the cavities.  (c) Weak coupling regime, with parameters $|g|/\kappa_{A}=|g|/\kappa_{B}=0.25, \Delta_{1}/\kappa_{A}=\Delta_{2}/\kappa_{B}=0.5$. This plot's differences from Fig.[$\ref{Fig2}$] are due to the presence of the second cavity and  are more clearly present in the weak coupling regime than in the strong coupling regime.}\label{Fig5}
\end{figure*}

Constructing the density operator as before, we can calculate the probabilities of finding the single excitation in different parts of our system as functions of time.
In Fig[\ref{Fig5}] we have plotted these probabilities as a function of time in both strong and weak coupling regimes. 
Like before, the single-photon Rabi oscillations and purely decaying behavior are clearly visible in the strong and weak coupling regimes, respectively. The new feature is a delayed probability of finding the photon in the second cavity as compared to the first cavity. This delay is now purely coming from the time the photon remains trapped in the cavities, as we have already eliminated the (trivial) delay between the cavities. How much that delay is,  and how the delay can be manipulated (increased, in particular) is discussed in the next subsection.

\subsection{Trapping of Photon and Time Dependent Spectra}
The simple cascaded system of two cavities can be used for slowing down the photon by trapping it in the second cavity \cite{khurgin2008slow,otey2008completely}. Looking at the maximum heights of the cavity probabilities in Fig.~(\ref{Fig5}b) already indicates that in the strong coupling regime, the photon is  delayed by a time of $\sim0.24\kappa_{1}^{-1}$ before reaching the second cavity. In the weak coupling regime this delay increases and approaches $\sim1.5\kappa_{1}^{-1}$. Moreover, we can increase the photon trapping time in the second cavity by setting the resonant frequency of the second cavity equal to one of the resonances visible in the spectrum  emitted by the first atom-cavity system. Analogous to Fig.~\ref{Fig5}(a)  we can plot the probabilities of finding the photon in the case $\Delta_{2}/\kappa_{2}=7.32$ corresponding to the right peak of the spectrum emitted by the first cavity (see Fig.[\ref{Fig3}-(a)]). By making that plot we can see that the strong coupling between the photon and the second cavity causes an extra delay  time of about $\sim 10\kappa_{1}^{-1}$, as if the photon is circulating many times before being reemitted into the fiber.

This delay can be further verified by looking at the time dependent spectrum \cite{eberly1977time} detected by detector $D_a$, as well as the synthesized version (i.e., the time-integrated version).
 \begin{figure*}[h]
\begin{center}
\begin{tabular}{cccc}
\subfloat{\includegraphics[width=7cm,height=5cm]{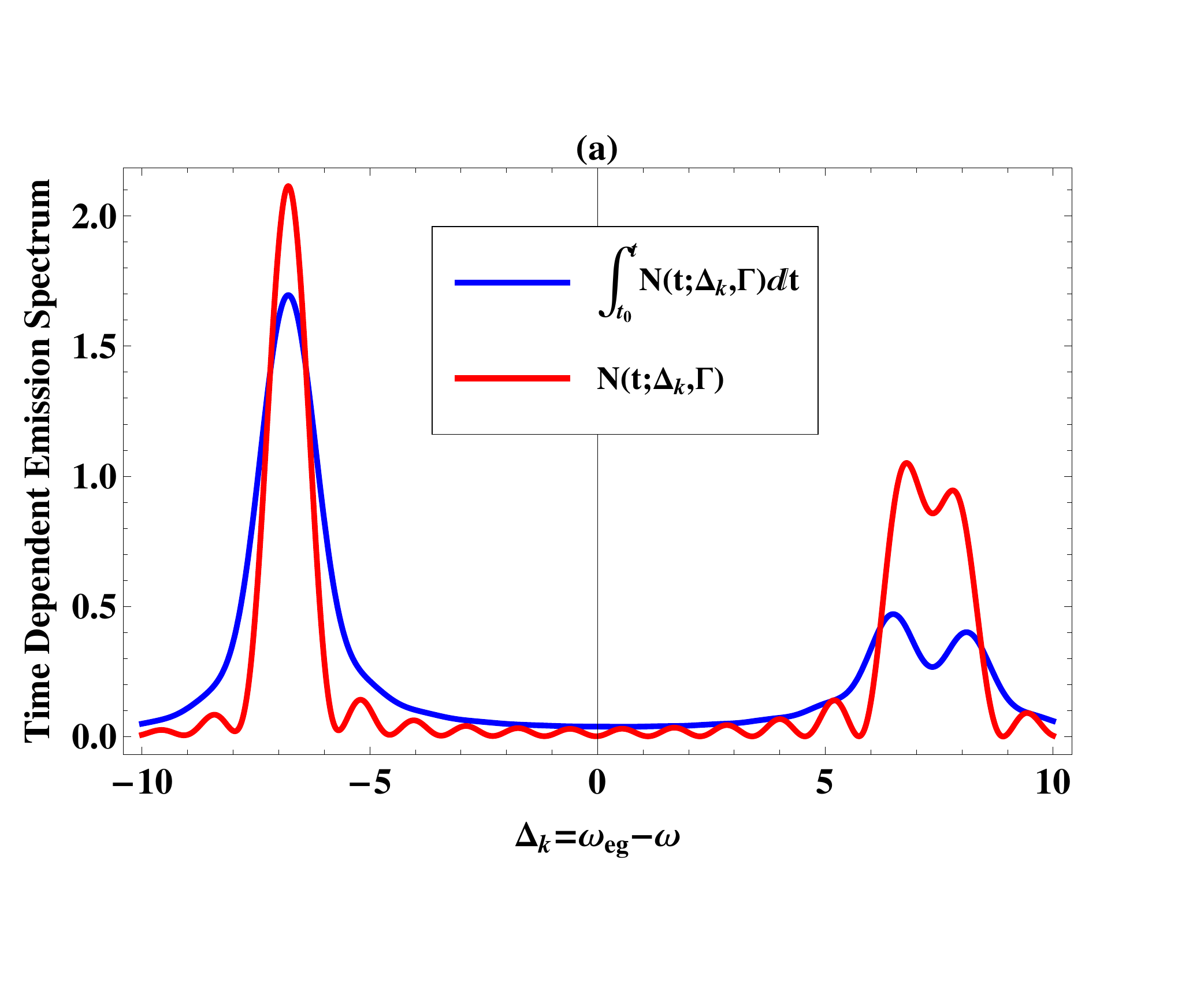}} & 
\subfloat{\includegraphics[width=7cm,height=5cm]{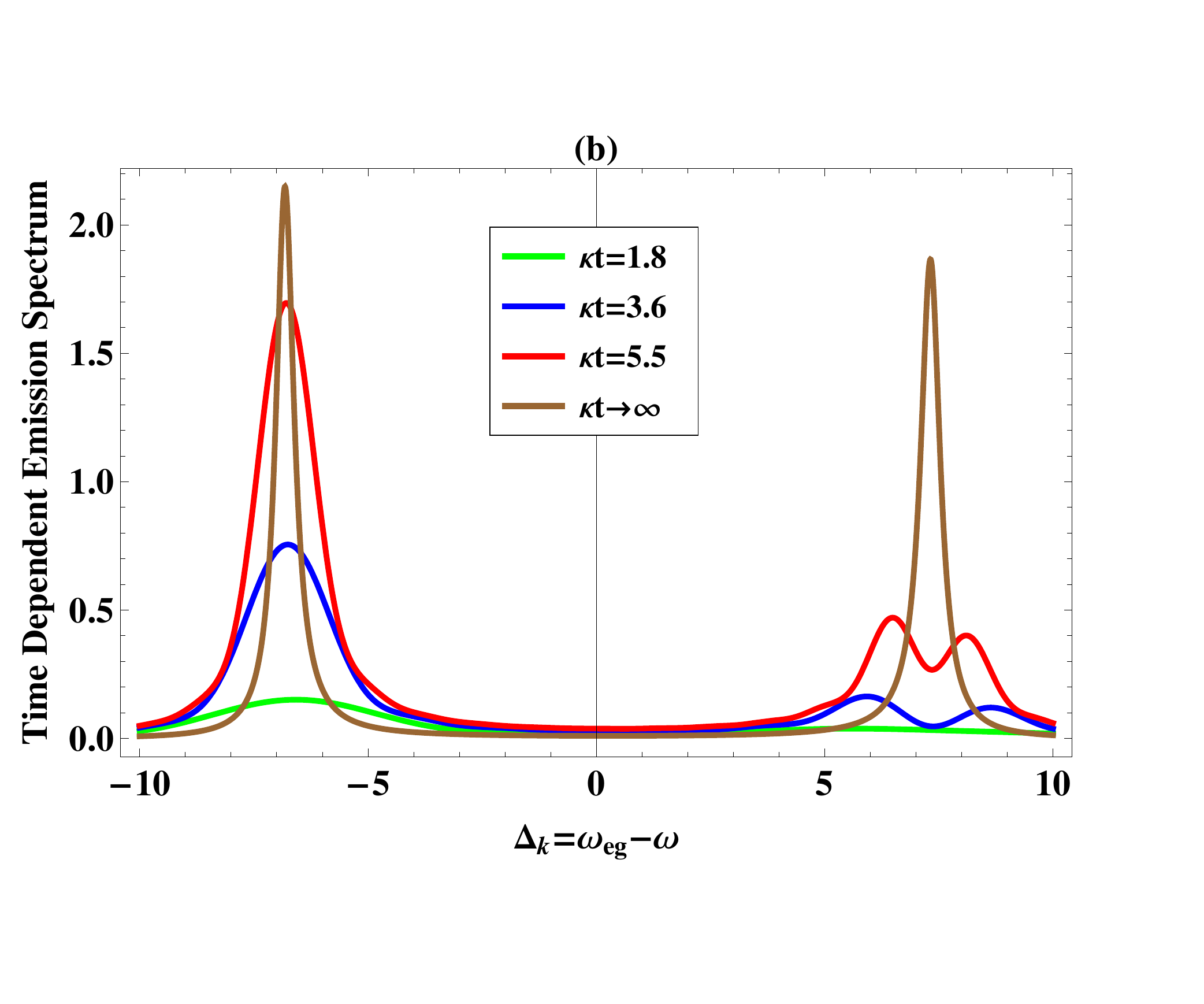}}\\
\end{tabular}
\captionsetup{
  margin=0em,
  justification=raggedright,
  singlelinecheck=false
}\caption{
(a) Comparison of the time-dependent spectrum as defined by either Eq.~(\ref{TDPS}) or (\ref{SPS}),  recorded at $\kappa t=5.5$ for $|g|/\kappa_{1}=|g|/\kappa_{2}=5$, $\Delta_{1}/\kappa_{1}=0.5$, $\Delta_{2}/\kappa_{2}=7.32$. Notice the role of integration is just to smooth out wiggles and changing the scale of the plot a little. Other features (in which we are more interested) remain the same. We now focus on the time-integrated version of the spectrum. (b) Integrated time dependent spectra recorded at different times for the same parameters as in (a), with  $\Gamma/\kappa_{1}=0.25$. Note that the peak on the right does not show considerable growth until $\kappa_{1}t=3.6$. In fact, we also see a ``hole-burning'' effect at earlier times for positive $\Delta_k$: there is one peak, but with the center frequencies removed (delayed).}\label{Fig6}
\end{center}
\end{figure*}
In Figs.~[\ref{Fig6}-(a)] we compare both types of time dependent spectra at  time $\kappa t=5.5$. We can see that the effect of the time integration is mainly to average out ripples, but apart from that the physical features we are interested in (which are discussed below) remain the same. From now on, we focus on the time-integrated spectrum as defined in (\ref{SPS}).

In Fig.~\ref{Fig6}-(b) we have plotted the time dependent spectra recorded by detector $D_a$ at $\kappa t=1.8, 3.6, 5.5$. We can see that at these times the left peak of the wave packet emitted by first cavity starts growing while the second peak is not considerably emitted until $\kappa t=3.6$. Even after that there is a small probability of finding a  photon recorded by the detector at $\Delta_{k} t=7.32$ which is due to the fact that photon remains trapped in the second cavity for a longer time before being emitted at this frequency. In the limit $t\rightarrow\infty$ we recover the single atom-cavity spectrum as plotted in Fig.[\ref{Fig3}]: the second cavity being linear does not change that spectrum.

\section{Array of ring resonators driven by one atom-cavity system}
\begin{figure*}[h]
  \begin{center}
\includegraphics[width=6in,height=1.7in]{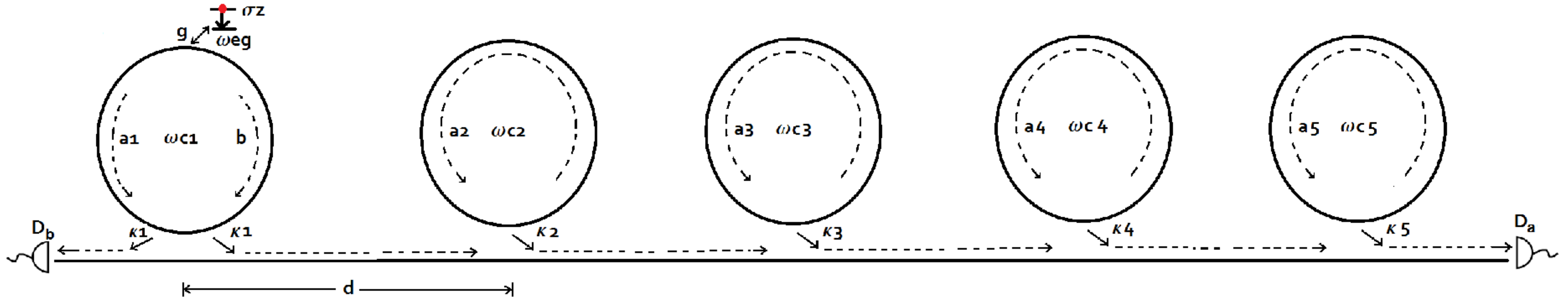}
\captionsetup{
  margin=0em,
  justification=raggedright,
  singlelinecheck=false
}  \caption{Array of four ring resonators driven by ring-atom system.}\label{Fig7}
  \end{center}
 \end{figure*}

\begin{figure*}[h]
\begin{center}
\begin{tabular}{cccc}
\subfloat{\includegraphics[width=6cm,height=5cm]{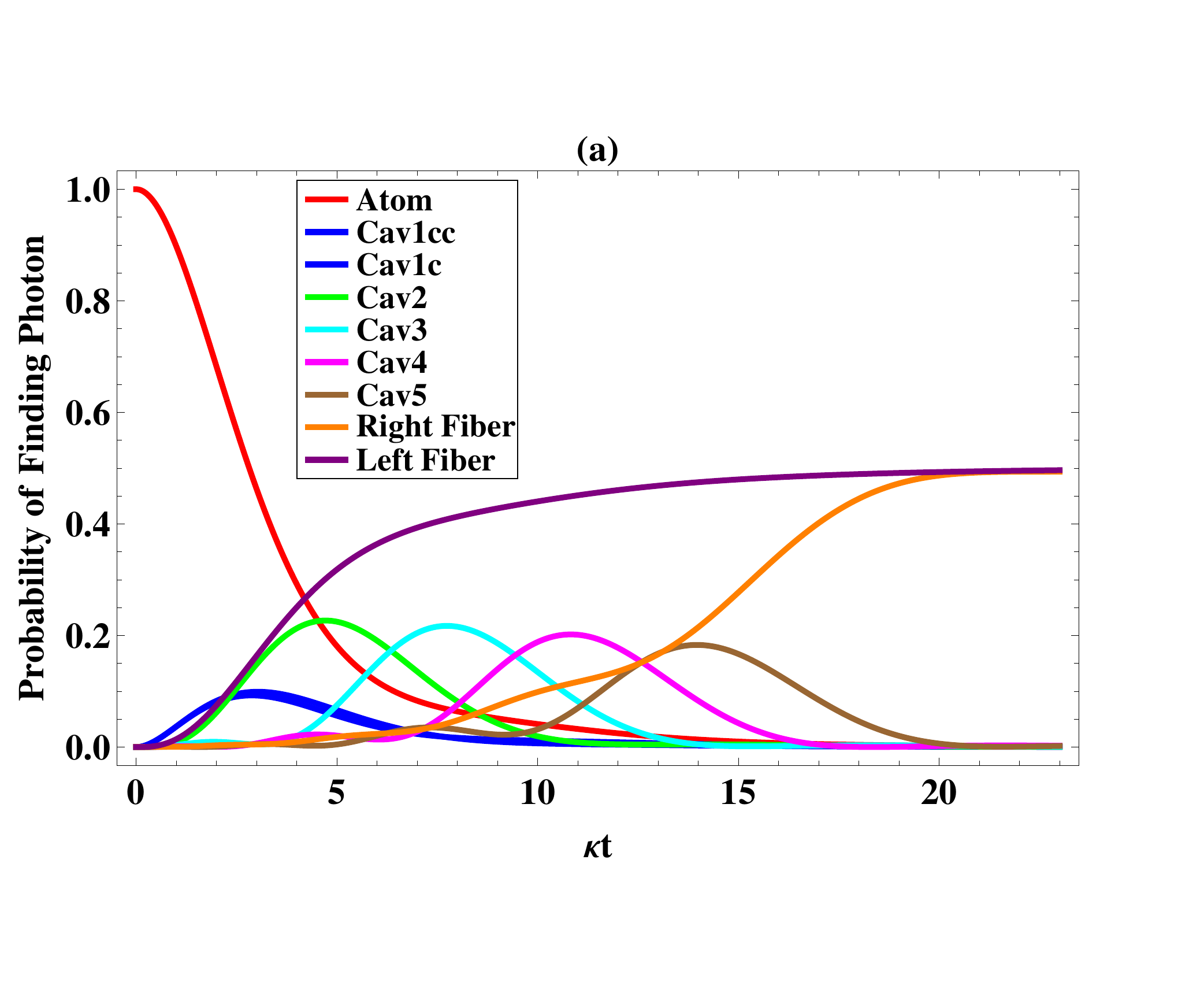}}&
\subfloat{\includegraphics[width=6cm,height=5cm]{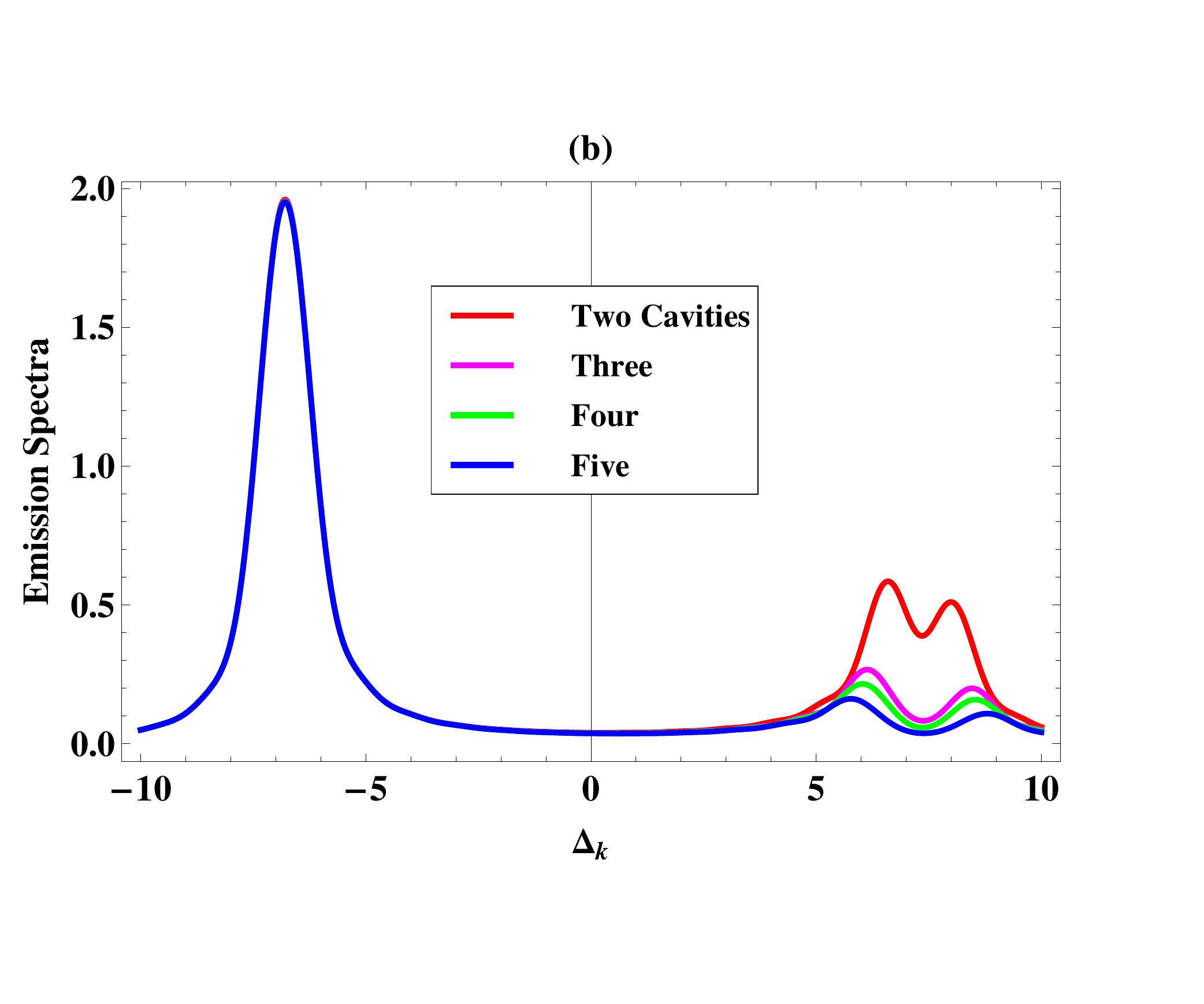}}\\
\end{tabular}
\captionsetup{
  margin=0em,
  justification=raggedright,
  singlelinecheck=false
}
\caption{
 (a) Probabilities of finding photon in atom, 1st cavity clockwise-anticlockwise modes, 2nd-5th cavity and in left-right fibers in the weak coupling regime $|g|/\kappa_{1}=...=|g|/\kappa_{5}=0.25, \Delta_{1}/\kappa_{1}=0.5$, $\Delta_{2}/\kappa_{2}=...=\Delta_{5}/\kappa_{5}=-0.12$. The successive delays in finding the photon in the different cavities are clearly visible. These delays are caused by the photon being trapped in each of the cavities for some time, and is in addition to the trivial delays caused by the propagation time between the cavities. In both regimes the photon can be trapped for times $\sim 15\kappa_1^{-1}$ in total. (b) Time-integrated spectra (recorded at $\kappa_1t=6$ with a detector with spectral width $\Gamma=0.25\kappa_1$) detected by $D_a$ detector in the strong coupling regime $|g|/\kappa_{1}=...=|g|/\kappa_{5}=5, \Delta_{1}/\kappa_{1}=0.5$, $\Delta_{2}/\kappa_{2}=...=\Delta_{5}/\kappa_{5}=7.32$.  Emission spectra are shown for two, three, four and five cavities. Note the ``hole-burning'' effect in the right peak in the spectrum. It is caused by the delay of the central frequency components of that peak.}\label{Fig8}
\end{center}
\end{figure*}

Our calculation can easily be generalized now to many cavities coupled to a single atom-cavity system. In the present section we are going to take the example of {\em four}  empty ring resonators driven by a single two-level excited atom coupled to a cavity, as shown in Fig.~[\ref{Fig7}].
Generalizing the procedure introduced in Section II  to four cavities, the non-Hermitian Hamiltonian in this example turns out to be
 \begin{equation}
 \begin{split}
 \hat{H}_{NH}
 &=-\hbar\omega_{eg}\hat{\sigma}_{-}\hat{\sigma}_{+}
 +\hbar\omega_{c}\hat{b}^{\dagger}\hat{b} +\hbar(g\hat{a}_{1}^{\dagger}\hat{\sigma}_{-} +g^{\ast}\hat{a}_{1}\hat{\sigma}_{+})\\
 &
 +\hbar(g^{\ast}\hat{b}^{\dagger}\hat{\sigma}_{-} +g\hat{b}\hat{\sigma}_{+}) +\sum_{i=1}^5\hbar(\omega_{ci}-\frac{i\kappa_{i}}{2})\hat{a}_{i}^{\dagger}\hat{a}_{i}\\
 &-i\hbar\sum_{i=1}^{4}\sum_{j=i+1}^5
 \sqrt{\kappa_{i}\kappa_{j}}
 \hat{a}_{j}^{\dagger}\hat{a}_{i}.
 \end{split}
\end{equation}
 Like before,  the jump operator corresponding to detections by detector $D_b$ is $\hat{J}_b=\hat{b}_{{\rm out}}=\sqrt{\kappa_{1}}\hat{b}$, and for $\hat{J}_a$ we simply generalize our previous result and obtain
\begin{equation}
\hat{J}_a=\hat{a}_{{\rm out}}^{(5)}=\sum_{j=2}^5\sqrt{\kappa_{j}}\hat{a}_{j}. 
\end{equation} 
Like before, we have absorbed spatial delays between cavities in the time arguments of the operators. Notice that now, from the above Hamiltonian (and in all cases of more than 2 cavities in array) the unidirectional coupling is such that a photon can be created in the last cavity by destroying a photon in any of the previous cavities. In Fig[\ref{Fig8}-(a)] we have plotted the probabilities of finding the excitation in the various cavities, in both strong and weak coupling regimes. 
The resonant frequency of all empty cavities in the array is chosen to be one of the peak frequencies of the spectrum emitted by the atom-cavity system so that the photon will remain trapped in each of the remaining four cavities, thus leading to a substantial delay. We see that by making this choice and taking $\kappa_{i}\equiv\kappa$ with $i=1,2,...,5$ we can trap the photon in cavities for more than a time $15\kappa^{-1}$  in both the strong  and weak coupling regimes. 

This trapping was further confirmed by looking at the time dependent spectra emitted by two, three, four and five cavities at $\kappa t=6$ as shown in Fig[\ref{Fig8}-(b)]. We can see that the probability of the photon being detected at $\kappa t=6$ is five times reduced in the case of five cavities compared to double cavity case. This indicates that, for the present case of five cavities, the photon remains trapped in the cavities for five times longer than in comparison to the two-cavities case discussed in Section II, which is consistent with our time evolution probability plot Fig[\ref{Fig8}-(a)].

\begin{figure}[h]
\includegraphics[width=7cm,height=5cm]{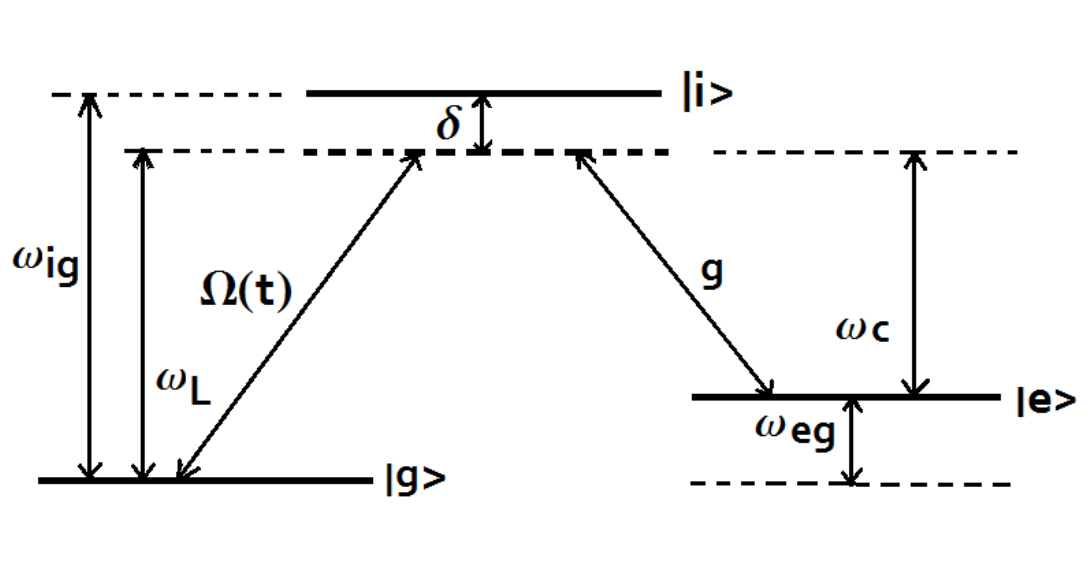}
\captionsetup{
  margin=0em,
  justification=raggedright,
  singlelinecheck=false
}
\caption{Raman type transition driven by laser and cavity. The atom starts in the state $\ket{g}$, and may be (with some probability between 0 and 1) transferred to the state $\ket{e}$ by absorbing a laser photon and emitting a photon into the cavity.}\label{Fig9a}
\end{figure}
\section{Replacing the two-level transition with Raman $\Lambda$ transition }
In practice, quantum information processing using atoms is often performed with a Raman transition between two ground states through a virtual intermediate excited state. This avoids spontaneous emission and guarantees long lifetimes of the two relevant ground states that store the quantum information.
In this section we will consider how the results of preceding sections are modified by replacing the two-level transition with a Raman transition. 
 
We examine, then, the system displayed in Fig.[\ref{Fig9a}], a three level atom in a $\Lambda$ configuration, with a ground state $\ket{g}$, a target/excited state $\ket{e}$ and an intermediate state $\ket{i}$ with energies $\hbar\omega_{g},\hbar\omega_{e}$ and $\hbar\omega_{i}$ respectively. Due to absence of direct coupling between ground and excited state, the far detuned intermediate state is used as a coupling route between these state. The transition from the ground to intermediate state is driven by a laser field with frequency $\omega_{L}$ and Rabi frequency $\Omega$ while the detuning is $\delta=\omega_{i}-\omega_{g}-\omega_{L}$. The transition from the intermediate state to the target state is carried out through  the coupling of the atom to the cavity mode, described by a coupling constant $g$. The
 Hamiltonian of such a laser-atom-cavity system is expressed as
\begin{equation}
\begin{split}
\hat{H}&=\hbar\omega_{eg}\ket{e}\bra{e}+\hbar\omega_{ig}\ket{i}\bra{i}+\hbar\omega_{c1}\hat{a}_{1}^{\dagger}\hat{a}_{1}+
\hbar\omega_{c1}\hat{b}^{\dagger}\hat{b}\\
&+\hbar\omega_{c2}\hat{a}_{2}^{\dagger}\hat{a}_{2}+
\frac{\hbar}{2}(\Omega(t) e^{i\omega_{L}t}\ket{g}\bra{i}+\Omega^{\ast}(t) e^{-i\omega_{L}t}\ket{i}\bra{g})\\
&+\hbar (g\hat{a}_{1}\ket{i}\bra{e}+g^{\ast}\hat{a}_{1}^{\dagger}\ket{e}\bra{i})+\hbar (g^{\ast}\hat{b}\ket{i}\bra{e}+g\hat{b}^{\dagger}\ket{e}\bra{i}),
\end{split}
\end{equation}
where $\omega_{e}-\omega_{g}\equiv\omega_{eg},\omega_{i}-\omega_{g}\equiv\omega_{ig}$ and the Rabi frequency $\Omega(t)$ can generally be time dependent. Going to a frame rotating with the laser frequency and adiabatically eliminating the intermediate state, the above described three-level system becomes an effective two-level system with  Hamiltonian
\begin{equation}
\begin{split}
\hat{H}&=\hbar\Delta_{c1}\hat{a}_{1}^{\dagger}\hat{a}_{1}+\hbar\Delta_{c1}\hat{b}^{\dagger}\hat{b}+\hbar\Delta_{c2}\hat{a}_{2}^{\dagger}\hat{a}_{2}
-\frac{\hbar|\Omega(t)|^2}{4\delta}\ket{g}\bra{g}\\
&-\frac{\hbar|g|^2}{\delta}\hat{a}_{1}^{\dagger}\hat{a}_{1}\ket{e}\bra{e}-\frac{\hbar|g|^2}{\delta}\hat{b}^{\dagger}\hat{b}\ket{e}\bra{e}\\
&-\hbar(\frac{g^{\ast}\Omega(t)}{2\delta}\ket{g}\bra{e}\hat{a}_{1}
+\frac{g\Omega(t)^{\ast}}{2\delta}\ket{e}\bra{g}\hat{a}_{1}^{\dagger})\\
&-\hbar(\frac{g\Omega(t)^{\ast}}{2\delta}\ket{g}\bra{e}\hat{b}
+\frac{g^{\ast}\Omega(t)}{2\delta}\ket{e}\bra{g}\hat{b}^{\dagger}),
\end{split}
\end{equation}
where $\Delta_{ci}\equiv\omega_{ci}-\omega_{L}$ for $i=1,2$, and $\frac{g^{\ast}\Omega(t)}{2\delta}$ can be thought as an effective coupling between the two-level atom and the cavity mode. 
This time-dependent coupling rate, appearing instead of the constant rate $g$, is one important difference with the case discussed in previous Sections. The other differences are the presence in the Hamiltonian of energy shifts (AC-Stark shifts)  of both states $\ket{e}$ and $\ket{g}$. Consequences of these differences are highlighted below. 

The derivation of the results follows the same lines, except that we did not obtain analytical results; instead we plot in the remaining 4 figures numerical results, both in the weak and strong coupling, and both time-dependent populations and time-dependent spectra. We assumed here a Gaussian function of time for the effective coupling rate, and one empty cavity driven by the three-level atom/cavity system (the extension to the case of multiple empty cavities is straightforward).

\begin{figure*}[h]
\begin{center}
\begin{tabular}{cccc}
\subfloat{\includegraphics[width=6cm,height=5cm]{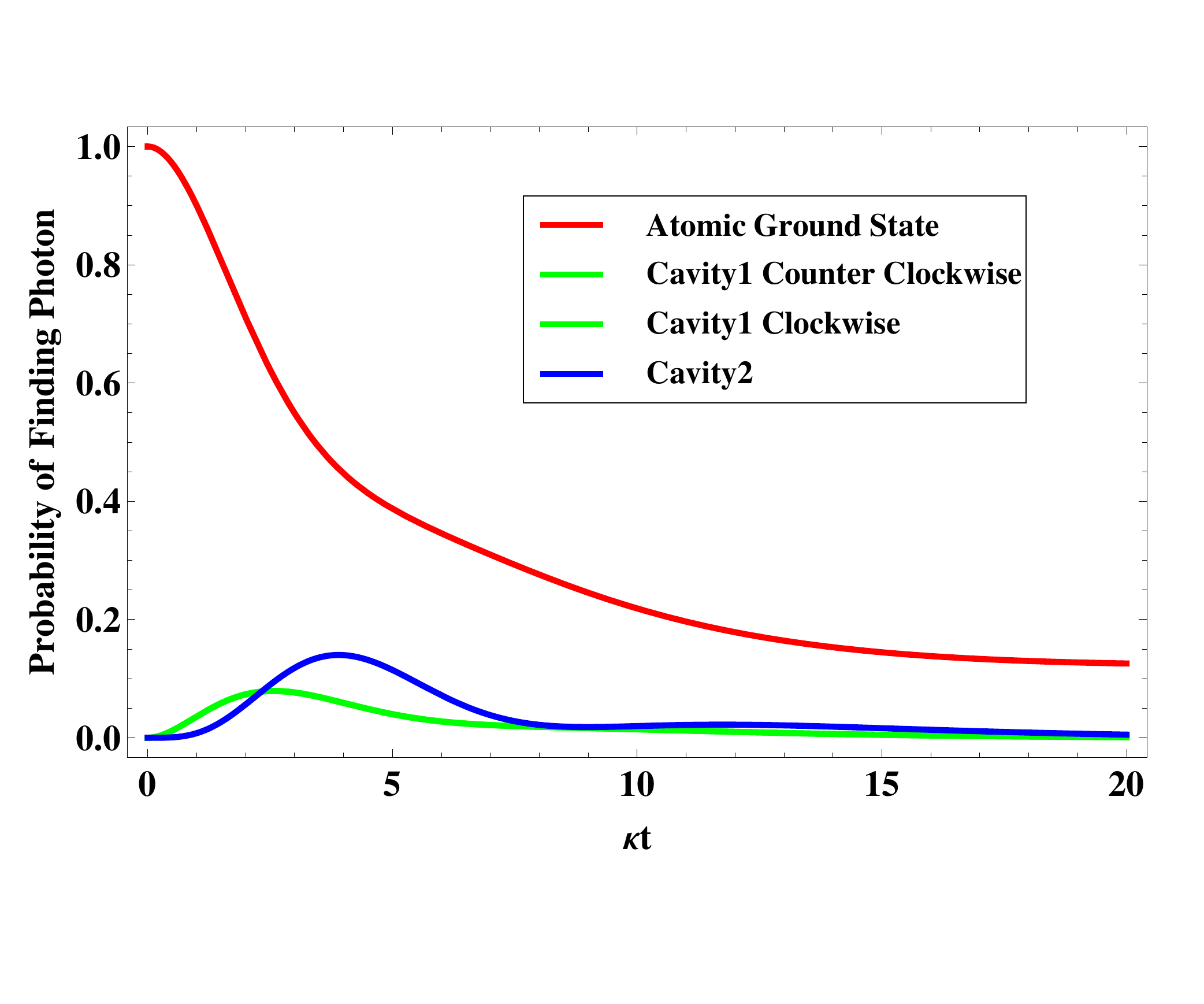}} & 
\subfloat{\includegraphics[width=6cm,height=5cm]{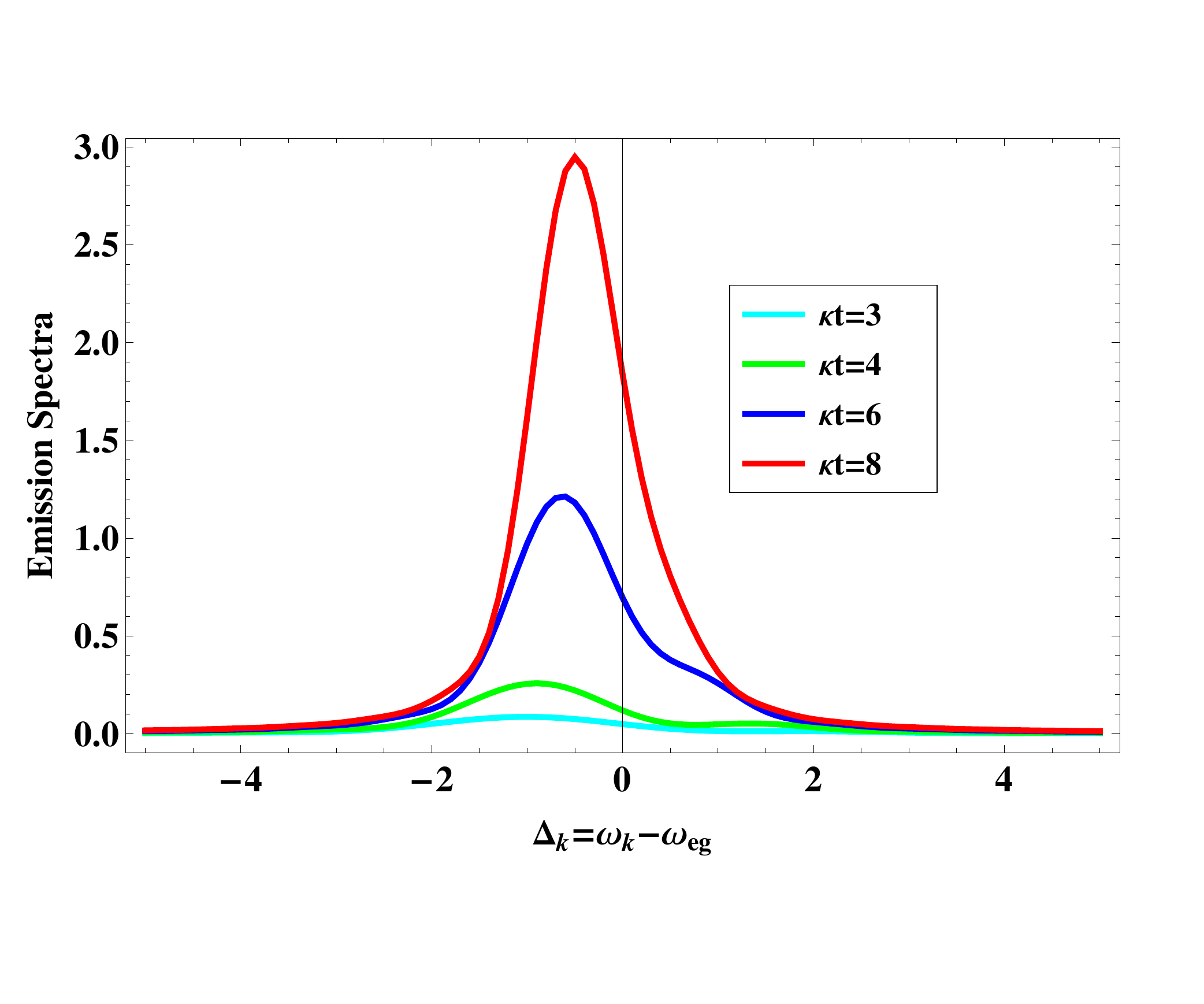}} \\
\end{tabular}  
\captionsetup{
  margin=0em,
  justification=raggedright,
  singlelinecheck=false
}\caption{(Figure on the left) Probabilities of finding the atom in the ground state, or an excitation  in the first or second cavity, assuming a Gaussian laser pulse with the form $\exp(-t^2/2\tau_L^2)$, in the weak coupling regime, with parameters $|g|/\kappa=0.25,\Delta_{c1}/\kappa=\Delta_{c2}/\kappa=0.25,\delta/\kappa=0.5$ and $\tau_L=10/\kappa$.
(Figure on the right) Integrated time- dependent spectra (recorded with bandwidth $\Gamma/\kappa_1=0.25$) emitted by cavity driven by Raman atom-cavity system recorded at different times for the same parameters.}\label{Fig9c}
\end{center}
\end{figure*}
\begin{figure*}[h]
\begin{center}
\begin{tabular}{cccc}
\subfloat{\includegraphics[width=6cm,height=5cm]{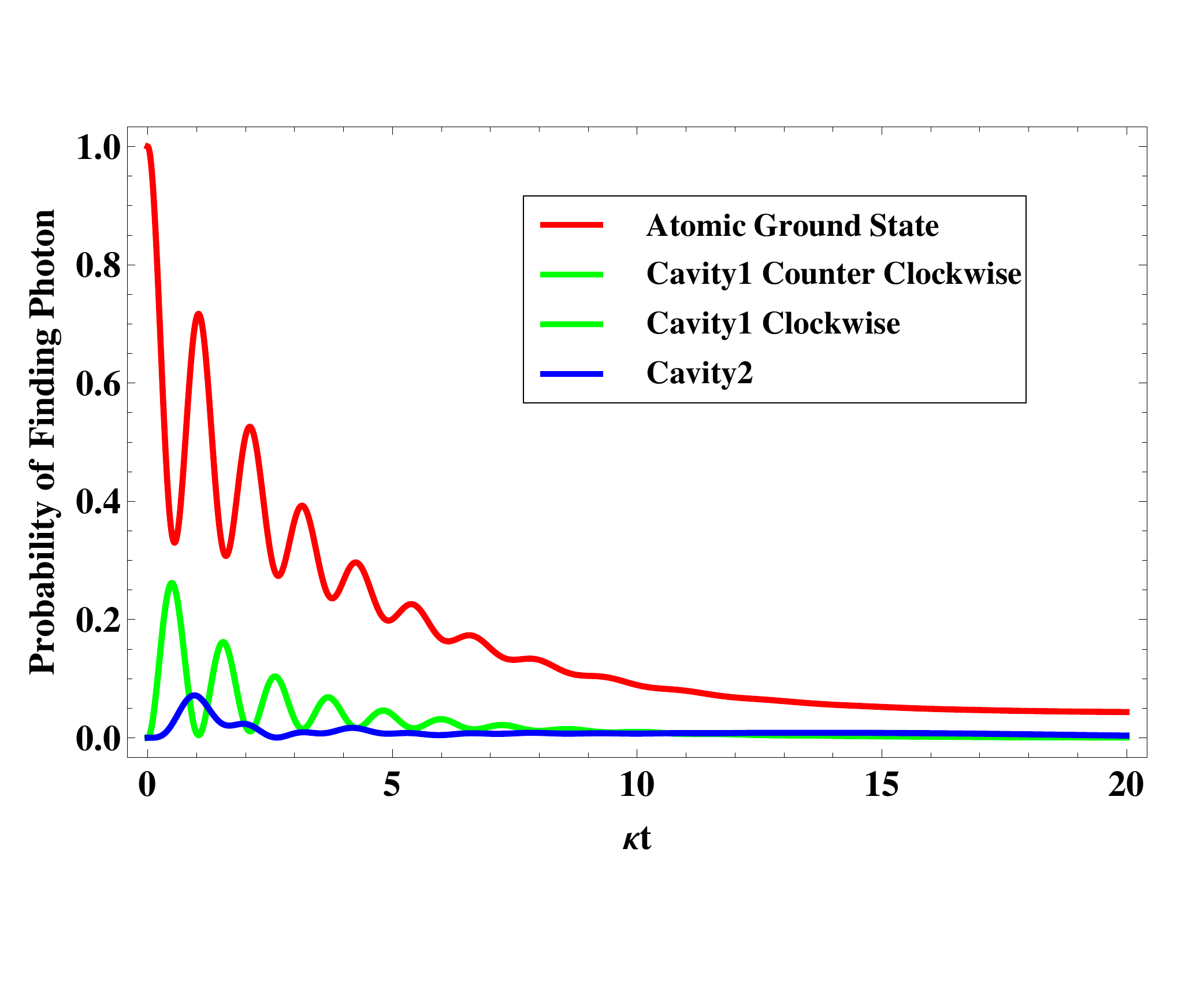}} & 
\subfloat{\includegraphics[width=6cm,height=5cm]{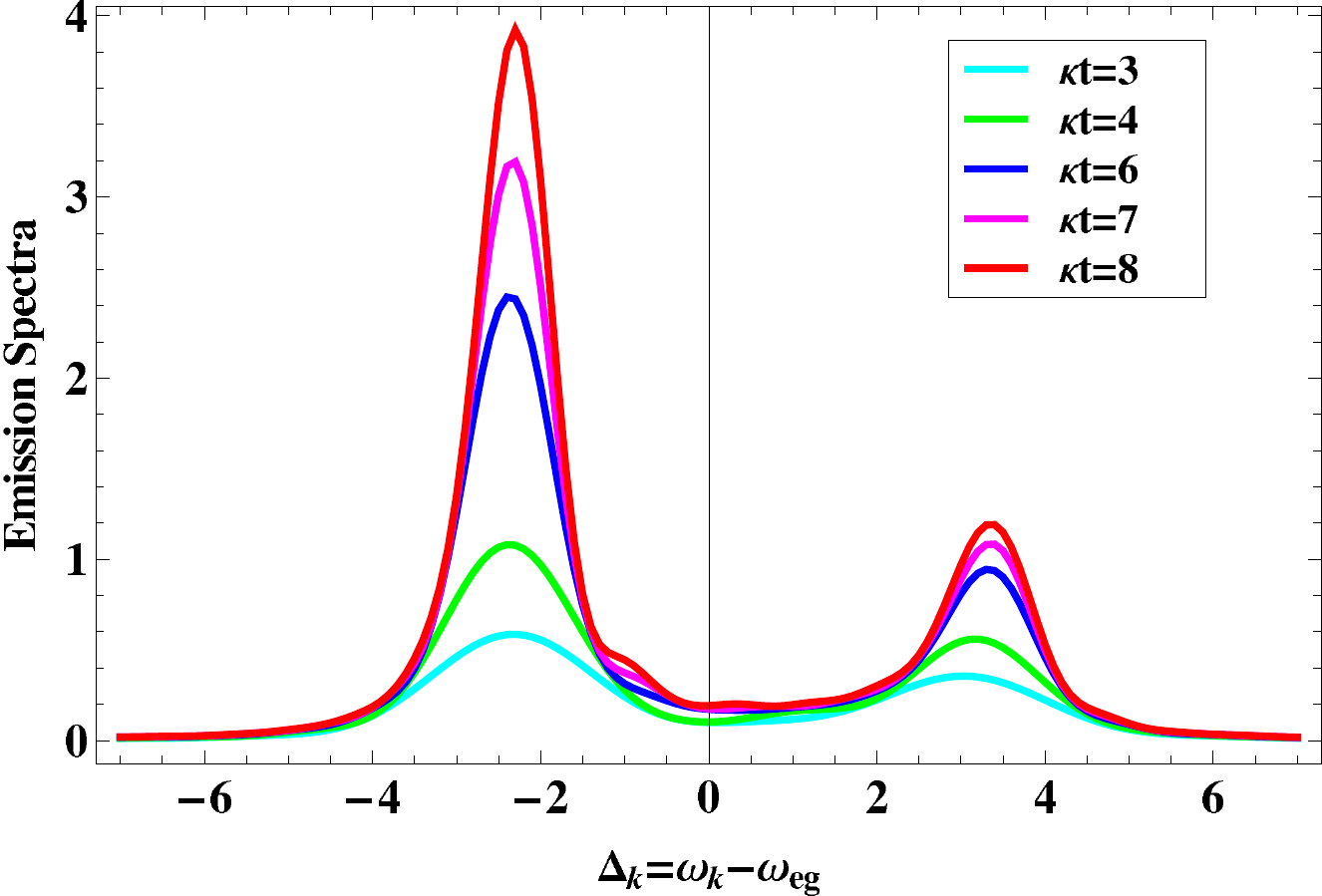}} \\
\end{tabular}
\captionsetup{
  margin=0em,
  justification=raggedright,
  singlelinecheck=false
}\caption{(Figure on the left) Probabilities of finding the atom in the ground state, or an excitation in the first or second cavity, in the strong coupling regime with $|g|/\kappa=2,\Delta_{c1}/\kappa=\Delta_{c2}/\kappa=0.25,\delta/\kappa=1.5$.
(Figure on the right) Integrated time-dependent spectra in the strong coupling regime, for the same parameters, with $\Gamma=0.25\kappa$. We note that changing the detunings ($\Delta_{c1}, \delta$) from positive to negative values would shift the graph towards the left; moreover, the heights of left and right peaks would be interchanged.}\label{Fig9e}
\end{center}
\end{figure*}

For the populations we see essentially the same sort of behaviour as we saw earlier,  except  that now the probabilities of finding the excitation in the various modes or the atom do not add up to unity because the laser does not necessarily succeed in creating an excitation. 

For the time-dependent spectra, too, we see more or less the same sort of behavior, except that the locations of the peaks are shifted thanks to the above-mentioned AC-Stark shifts.

\section{Conclusions}
The main novelty in our paper is the calculation of time-dependent spectra of single photons propagating through coupled cavity arrays. Our calculations take into account how the single photon is produced (by an atom or a quantum dot or a NV center in diamond inside one resonator), and how it subsequently  travels through the remaining empty cavities before being detected.
We found that the delay of frequency components resonant with one or more empty cavities is nicely represented in the time-dependent spectrum.
We see, for example, ``hole-burning'' effects where at earlier times a broad peak appears with a hole at those frequencies that are delayed (and which show up in the later spectra).
\section*{Acknowledgments}
HJK acknowledges funding from the IQIM, an NSF Physics Frontier Center with support of 
the Moore Foundation, by the AFOSR QuMPASS MURI, by the DoD NSSEFF program, and by NSF PHY-1205729.
\bibliography{Article4}

\begin{thebibliography}{36}
\expandafter\ifx\csname natexlab\endcsname\relax\def\natexlab#1{#1}\fi
\expandafter\ifx\csname bibnamefont\endcsname\relax
  \def\bibnamefont#1{#1}\fi
\expandafter\ifx\csname bibfnamefont\endcsname\relax
  \def\bibfnamefont#1{#1}\fi
\expandafter\ifx\csname citenamefont\endcsname\relax
  \def\citenamefont#1{#1}\fi
\expandafter\ifx\csname url\endcsname\relax
  \def\url#1{\texttt{#1}}\fi
\expandafter\ifx\csname urlprefix\endcsname\relax\def\urlprefix{URL }\fi
\providecommand{\bibinfo}[2]{#2}
\providecommand{\eprint}[2][]{\url{#2}}

\bibitem[{\citenamefont{Yariv et~al.}(1999)\citenamefont{Yariv, Xu, Lee, and
  Scherer}}]{yariv1999coupled}
\bibinfo{author}{\bibfnamefont{A.}~\bibnamefont{Yariv}},
  \bibinfo{author}{\bibfnamefont{Y.}~\bibnamefont{Xu}},
  \bibinfo{author}{\bibfnamefont{R.}~\bibnamefont{Lee}}, \bibnamefont{and}
  \bibinfo{author}{\bibfnamefont{A.}~\bibnamefont{Scherer}},
  \bibinfo{journal}{Optics Lett.} \textbf{\bibinfo{volume}{24}},
  \bibinfo{pages}{711} (\bibinfo{year}{1999}).

\bibitem[{\citenamefont{Heebner et~al.}(2002)\citenamefont{Heebner, Boyd, and
  Park}}]{heebner2002scissor}
\bibinfo{author}{\bibfnamefont{J.}~\bibnamefont{Heebner}},
  \bibinfo{author}{\bibfnamefont{R.}~\bibnamefont{Boyd}}, \bibnamefont{and}
  \bibinfo{author}{\bibfnamefont{Q.}~\bibnamefont{Park}},
  \bibinfo{journal}{JOSA B} \textbf{\bibinfo{volume}{19}}, \bibinfo{pages}{722}
  (\bibinfo{year}{2002}).

\bibitem[{\citenamefont{Heebner and Boyd}(2002)}]{heebner2002slow}
\bibinfo{author}{\bibfnamefont{J.}~\bibnamefont{Heebner}} \bibnamefont{and}
  \bibinfo{author}{\bibfnamefont{R.}~\bibnamefont{Boyd}}, \bibinfo{journal}{J.
  Mod. Opt.} \textbf{\bibinfo{volume}{49}}, \bibinfo{pages}{2629}
  (\bibinfo{year}{2002}).

\bibitem[{\citenamefont{Scheuer et~al.}(2005)\citenamefont{Scheuer, Paloczi,
  Poon, and Yariv}}]{scheuer2005coupled}
\bibinfo{author}{\bibfnamefont{J.}~\bibnamefont{Scheuer}},
  \bibinfo{author}{\bibfnamefont{G.}~\bibnamefont{Paloczi}},
  \bibinfo{author}{\bibfnamefont{J.}~\bibnamefont{Poon}}, \bibnamefont{and}
  \bibinfo{author}{\bibfnamefont{A.}~\bibnamefont{Yariv}},
  \bibinfo{journal}{Optics and photonics news} \textbf{\bibinfo{volume}{16}},
  \bibinfo{pages}{36} (\bibinfo{year}{2005}).

\bibitem[{\citenamefont{Baba}(2008)}]{baba2008slow}
\bibinfo{author}{\bibfnamefont{T.}~\bibnamefont{Baba}},
  \bibinfo{journal}{Nature Photonics} \textbf{\bibinfo{volume}{2}},
  \bibinfo{pages}{465} (\bibinfo{year}{2008}).

\bibitem[{\citenamefont{Krauss}(2008)}]{krauss2008we}
\bibinfo{author}{\bibfnamefont{T.}~\bibnamefont{Krauss}},
  \bibinfo{journal}{Nature Photonics} \textbf{\bibinfo{volume}{2}},
  \bibinfo{pages}{448} (\bibinfo{year}{2008}).

\bibitem[{\citenamefont{Shen and Fan}(2005)}]{shen2005coherent}
\bibinfo{author}{\bibfnamefont{J.}~\bibnamefont{Shen}} \bibnamefont{and}
  \bibinfo{author}{\bibfnamefont{S.}~\bibnamefont{Fan}},
  \bibinfo{journal}{Optics Lett.} \textbf{\bibinfo{volume}{30}},
  \bibinfo{pages}{2001} (\bibinfo{year}{2005}).

\bibitem[{\citenamefont{Shen and Fan}(2009{\natexlab{a}})}]{shen2009theory1}
\bibinfo{author}{\bibfnamefont{J.}~\bibnamefont{Shen}} \bibnamefont{and}
  \bibinfo{author}{\bibfnamefont{S.}~\bibnamefont{Fan}},
  \bibinfo{journal}{Phys. Rev. A} \textbf{\bibinfo{volume}{79}},
  \bibinfo{pages}{023837} (\bibinfo{year}{2009}{\natexlab{a}}).

\bibitem[{\citenamefont{Shen and Fan}(2009{\natexlab{b}})}]{shen2009theory}
\bibinfo{author}{\bibfnamefont{J.}~\bibnamefont{Shen}} \bibnamefont{and}
  \bibinfo{author}{\bibfnamefont{S.}~\bibnamefont{Fan}},
  \bibinfo{journal}{Phys. Rev. A} \textbf{\bibinfo{volume}{79}},
  \bibinfo{pages}{023838} (\bibinfo{year}{2009}{\natexlab{b}}).

\bibitem[{\citenamefont{Rephaeli and Fan}(2012)}]{rephaeli2012stimulated}
\bibinfo{author}{\bibfnamefont{E.}~\bibnamefont{Rephaeli}} \bibnamefont{and}
  \bibinfo{author}{\bibfnamefont{S.}~\bibnamefont{Fan}},
  \bibinfo{journal}{Phys. Rev. Lett.} \textbf{\bibinfo{volume}{108}},
  \bibinfo{pages}{143602} (\bibinfo{year}{2012}).

\bibitem[{\citenamefont{Santori et~al.}(2002)\citenamefont{Santori,
  David~Fattal, Solomon, and Yamamoto}}]{santori2002indistinguishable}
\bibinfo{author}{\bibfnamefont{C.}~\bibnamefont{Santori}},
  \bibinfo{author}{\bibfnamefont{J.}~\bibnamefont{David~Fattal}},
  \bibinfo{author}{\bibfnamefont{G.}~\bibnamefont{Solomon}}, \bibnamefont{and}
  \bibinfo{author}{\bibfnamefont{Y.}~\bibnamefont{Yamamoto}},
  \bibinfo{journal}{Nature} \textbf{\bibinfo{volume}{419}},
  \bibinfo{pages}{594} (\bibinfo{year}{2002}).

\bibitem[{\citenamefont{Kuhn et~al.}(2002)\citenamefont{Kuhn, Hennrich, and
  Rempe}}]{kuhn2002deterministic}
\bibinfo{author}{\bibfnamefont{A.}~\bibnamefont{Kuhn}},
  \bibinfo{author}{\bibfnamefont{M.}~\bibnamefont{Hennrich}}, \bibnamefont{and}
  \bibinfo{author}{\bibfnamefont{G.}~\bibnamefont{Rempe}},
  \bibinfo{journal}{Phys. Rev. Lett.} \textbf{\bibinfo{volume}{89}},
  \bibinfo{pages}{67901} (\bibinfo{year}{2002}).

\bibitem[{\citenamefont{McKeever et~al.}(2004)\citenamefont{McKeever, Boca,
  Boozer, Miller, Buck, Kuzmich, and Kimble}}]{mckeever2004deterministic}
\bibinfo{author}{\bibfnamefont{J.}~\bibnamefont{McKeever}},
  \bibinfo{author}{\bibfnamefont{A.}~\bibnamefont{Boca}},
  \bibinfo{author}{\bibfnamefont{A.}~\bibnamefont{Boozer}},
  \bibinfo{author}{\bibfnamefont{R.}~\bibnamefont{Miller}},
  \bibinfo{author}{\bibfnamefont{J.}~\bibnamefont{Buck}},
  \bibinfo{author}{\bibfnamefont{A.}~\bibnamefont{Kuzmich}}, \bibnamefont{and}
  \bibinfo{author}{\bibfnamefont{H.}~\bibnamefont{Kimble}},
  \bibinfo{journal}{Science} \textbf{\bibinfo{volume}{303}},
  \bibinfo{pages}{1992} (\bibinfo{year}{2004}).

\bibitem[{\citenamefont{Englund et~al.}(2010)\citenamefont{Englund, Shields,
  Rivoire, Hatami, Vuckovic, Park, and Lukin}}]{englund2010}
\bibinfo{author}{\bibfnamefont{D.}~\bibnamefont{Englund}},
  \bibinfo{author}{\bibfnamefont{B.}~\bibnamefont{Shields}},
  \bibinfo{author}{\bibfnamefont{K.}~\bibnamefont{Rivoire}},
  \bibinfo{author}{\bibfnamefont{F.}~\bibnamefont{Hatami}},
  \bibinfo{author}{\bibfnamefont{J.}~\bibnamefont{Vuckovic}},
  \bibinfo{author}{\bibfnamefont{H.}~\bibnamefont{Park}}, \bibnamefont{and}
  \bibinfo{author}{\bibfnamefont{M.~D.} \bibnamefont{Lukin}},
  \bibinfo{journal}{Nano Lett.} \textbf{\bibinfo{volume}{10}},
  \bibinfo{pages}{3922} (\bibinfo{year}{2010}).

\bibitem[{\citenamefont{Riedrich-M{\"o}ller
  et~al.}(2011)\citenamefont{Riedrich-M{\"o}ller, Kipfstuhl, Hepp, Neu, Pauly,
  M{\"u}cklich, Baur, Wandt, Wolff, Fischer et~al.}}]{riedrich2011}
\bibinfo{author}{\bibfnamefont{J.}~\bibnamefont{Riedrich-M{\"o}ller}},
  \bibinfo{author}{\bibfnamefont{L.}~\bibnamefont{Kipfstuhl}},
  \bibinfo{author}{\bibfnamefont{C.}~\bibnamefont{Hepp}},
  \bibinfo{author}{\bibfnamefont{E.}~\bibnamefont{Neu}},
  \bibinfo{author}{\bibfnamefont{C.}~\bibnamefont{Pauly}},
  \bibinfo{author}{\bibfnamefont{F.}~\bibnamefont{M{\"u}cklich}},
  \bibinfo{author}{\bibfnamefont{A.}~\bibnamefont{Baur}},
  \bibinfo{author}{\bibfnamefont{M.}~\bibnamefont{Wandt}},
  \bibinfo{author}{\bibfnamefont{S.}~\bibnamefont{Wolff}},
  \bibinfo{author}{\bibfnamefont{M.}~\bibnamefont{Fischer}},
  \bibnamefont{et~al.}, \bibinfo{journal}{Nature nanotechnology}
  \textbf{\bibinfo{volume}{7}}, \bibinfo{pages}{69} (\bibinfo{year}{2011}).

\bibitem[{\citenamefont{Carmichael}(1993{\natexlab{a}})}]{carmichael1993quantum}
\bibinfo{author}{\bibfnamefont{H.}~\bibnamefont{Carmichael}},
  \bibinfo{journal}{Phys. Rev. Lett.} \textbf{\bibinfo{volume}{70}},
  \bibinfo{pages}{2273} (\bibinfo{year}{1993}{\natexlab{a}}).

\bibitem[{\citenamefont{Gardiner}(1993)}]{gardiner1993driving}
\bibinfo{author}{\bibfnamefont{C.}~\bibnamefont{Gardiner}},
  \bibinfo{journal}{Phys. Rev. Lett.} \textbf{\bibinfo{volume}{70}},
  \bibinfo{pages}{2269} (\bibinfo{year}{1993}).

\bibitem[{\citenamefont{Gardiner and Zoller}(2004)}]{gardiner2004quantum}
\bibinfo{author}{\bibfnamefont{C.}~\bibnamefont{Gardiner}} \bibnamefont{and}
  \bibinfo{author}{\bibfnamefont{P.}~\bibnamefont{Zoller}},
  \emph{\bibinfo{title}{Quantum noise: a handbook of Markovian and
  non-Markovian quantum stochastic methods with applications to quantum
  optics}}, vol.~\bibinfo{volume}{56} (\bibinfo{publisher}{Springer},
  \bibinfo{year}{2004}).

\bibitem[{\citenamefont{Zoller et~al.}(1987)\citenamefont{Zoller, Marte, and
  Walls}}]{zoller1987}
\bibinfo{author}{\bibfnamefont{P.}~\bibnamefont{Zoller}},
  \bibinfo{author}{\bibfnamefont{M.}~\bibnamefont{Marte}}, \bibnamefont{and}
  \bibinfo{author}{\bibfnamefont{D.}~\bibnamefont{Walls}},
  \bibinfo{journal}{Physical Review A} \textbf{\bibinfo{volume}{35}},
  \bibinfo{pages}{198} (\bibinfo{year}{1987}).

\bibitem[{\citenamefont{Carmichael}(1993{\natexlab{b}})}]{carmichael1993open}
\bibinfo{author}{\bibfnamefont{H.}~\bibnamefont{Carmichael}},
  \emph{\bibinfo{title}{An open systems approach to Quantum Optics}}
  (\bibinfo{publisher}{Springer}, \bibinfo{year}{1993}{\natexlab{b}}).

\bibitem[{\citenamefont{Dum et~al.}(1992)\citenamefont{Dum, Parkins, Zoller,
  and Gardiner}}]{dum1992monte}
\bibinfo{author}{\bibfnamefont{R.}~\bibnamefont{Dum}},
  \bibinfo{author}{\bibfnamefont{A.}~\bibnamefont{Parkins}},
  \bibinfo{author}{\bibfnamefont{P.}~\bibnamefont{Zoller}}, \bibnamefont{and}
  \bibinfo{author}{\bibfnamefont{C.}~\bibnamefont{Gardiner}},
  \bibinfo{journal}{Phys. Rev. A} \textbf{\bibinfo{volume}{46}},
  \bibinfo{pages}{4382} (\bibinfo{year}{1992}).

\bibitem[{\citenamefont{M{\o}lmer et~al.}(1993)\citenamefont{M{\o}lmer, Castin,
  and Dalibard}}]{molmer1993monte}
\bibinfo{author}{\bibfnamefont{K.}~\bibnamefont{M{\o}lmer}},
  \bibinfo{author}{\bibfnamefont{Y.}~\bibnamefont{Castin}}, \bibnamefont{and}
  \bibinfo{author}{\bibfnamefont{J.}~\bibnamefont{Dalibard}},
  \bibinfo{journal}{J. Opt. Soc. Am. B} \textbf{\bibinfo{volume}{10}},
  \bibinfo{pages}{524} (\bibinfo{year}{1993}).

\bibitem[{\citenamefont{Plenio and Knight}(1998)}]{plenio1998quantum}
\bibinfo{author}{\bibfnamefont{M.}~\bibnamefont{Plenio}} \bibnamefont{and}
  \bibinfo{author}{\bibfnamefont{P.}~\bibnamefont{Knight}},
  \bibinfo{journal}{Rev. Mod. Phys.} \textbf{\bibinfo{volume}{70}},
  \bibinfo{pages}{101} (\bibinfo{year}{1998}).

\bibitem[{\citenamefont{Mandel}(1999)}]{mandel}
\bibinfo{author}{\bibfnamefont{L.}~\bibnamefont{Mandel}},
  \bibinfo{journal}{Rev. Mod. Phys.} \textbf{\bibinfo{volume}{71}},
  \bibinfo{pages}{S274} (\bibinfo{year}{1999}),
  \urlprefix\url{http://link.aps.org/doi/10.1103/RevModPhys.71.S274}.

\bibitem[{\citenamefont{Aspect et~al.}(1980)\citenamefont{Aspect, Roger,
  Reynaud, Dalibard, and Cohen-Tannoudji}}]{aspect1980}
\bibinfo{author}{\bibfnamefont{A.}~\bibnamefont{Aspect}},
  \bibinfo{author}{\bibfnamefont{G.}~\bibnamefont{Roger}},
  \bibinfo{author}{\bibfnamefont{S.}~\bibnamefont{Reynaud}},
  \bibinfo{author}{\bibfnamefont{J.}~\bibnamefont{Dalibard}}, \bibnamefont{and}
  \bibinfo{author}{\bibfnamefont{C.}~\bibnamefont{Cohen-Tannoudji}},
  \bibinfo{journal}{Physical Review Letters} \textbf{\bibinfo{volume}{45}},
  \bibinfo{pages}{617} (\bibinfo{year}{1980}).

\bibitem[{\citenamefont{Eberly and Wodkiewicz}(1977)}]{eberly1977time}
\bibinfo{author}{\bibfnamefont{J.}~\bibnamefont{Eberly}} \bibnamefont{and}
  \bibinfo{author}{\bibfnamefont{K.}~\bibnamefont{Wodkiewicz}},
  \bibinfo{journal}{J. Opt. Soc. Am.} \textbf{\bibinfo{volume}{67}},
  \bibinfo{pages}{1252} (\bibinfo{year}{1977}).

\bibitem[{\citenamefont{Dayan et~al.}(2008)\citenamefont{Dayan, Parkins, Aoki,
  Ostby, Vahala, and Kimble}}]{dayan2008photon}
\bibinfo{author}{\bibfnamefont{B.}~\bibnamefont{Dayan}},
  \bibinfo{author}{\bibfnamefont{A.}~\bibnamefont{Parkins}},
  \bibinfo{author}{\bibfnamefont{T.}~\bibnamefont{Aoki}},
  \bibinfo{author}{\bibfnamefont{E.}~\bibnamefont{Ostby}},
  \bibinfo{author}{\bibfnamefont{K.}~\bibnamefont{Vahala}}, \bibnamefont{and}
  \bibinfo{author}{\bibfnamefont{H.}~\bibnamefont{Kimble}},
  \bibinfo{journal}{Science} \textbf{\bibinfo{volume}{319}},
  \bibinfo{pages}{1062} (\bibinfo{year}{2008}).

\bibitem[{\citenamefont{Cui and Raymer}(2006)}]{cui2006emission}
\bibinfo{author}{\bibfnamefont{G.}~\bibnamefont{Cui}} \bibnamefont{and}
  \bibinfo{author}{\bibfnamefont{M.}~\bibnamefont{Raymer}},
  \bibinfo{journal}{Phys. Rev. A} \textbf{\bibinfo{volume}{73}},
  \bibinfo{pages}{053807} (\bibinfo{year}{2006}).

\bibitem[{\citenamefont{Auff{\`e}ves et~al.}(2008)\citenamefont{Auff{\`e}ves,
  Besga, G{\'e}rard, and Poizat}}]{auffeves2008spontaneous}
\bibinfo{author}{\bibfnamefont{A.}~\bibnamefont{Auff{\`e}ves}},
  \bibinfo{author}{\bibfnamefont{B.}~\bibnamefont{Besga}},
  \bibinfo{author}{\bibfnamefont{J.}~\bibnamefont{G{\'e}rard}},
  \bibnamefont{and} \bibinfo{author}{\bibfnamefont{J.}~\bibnamefont{Poizat}},
  \bibinfo{journal}{Phys. Rev. A} \textbf{\bibinfo{volume}{77}},
  \bibinfo{pages}{063833} (\bibinfo{year}{2008}).

\bibitem[{\citenamefont{Tian and Carmichael}(1992)}]{tian1992quantum}
\bibinfo{author}{\bibfnamefont{L.}~\bibnamefont{Tian}} \bibnamefont{and}
  \bibinfo{author}{\bibfnamefont{H.}~\bibnamefont{Carmichael}},
  \bibinfo{journal}{Phys. Rev. A} \textbf{\bibinfo{volume}{46}},
  \bibinfo{pages}{6801} (\bibinfo{year}{1992}).

\bibitem[{\citenamefont{Havukainen and Stenholm}(1998)}]{havukainen1998open}
\bibinfo{author}{\bibfnamefont{M.}~\bibnamefont{Havukainen}} \bibnamefont{and}
  \bibinfo{author}{\bibfnamefont{S.}~\bibnamefont{Stenholm}},
  \bibinfo{journal}{J. Mod. Opt.} \textbf{\bibinfo{volume}{45}},
  \bibinfo{pages}{1699} (\bibinfo{year}{1998}).

\bibitem[{\citenamefont{Carmichael}(2007)}]{carmichael2007statistical}
\bibinfo{author}{\bibfnamefont{H.}~\bibnamefont{Carmichael}},
  \emph{\bibinfo{title}{Statistical Methods in Quantum Optics, Vol.2}}
  (\bibinfo{publisher}{Springer}, \bibinfo{year}{2007}).

\bibitem[{\citenamefont{Carmichael et~al.}(1989)\citenamefont{Carmichael,
  Brecha, Raizen, Kimble, and Rice}}]{carmichael1989subnatural}
\bibinfo{author}{\bibfnamefont{H.}~\bibnamefont{Carmichael}},
  \bibinfo{author}{\bibfnamefont{R.}~\bibnamefont{Brecha}},
  \bibinfo{author}{\bibfnamefont{M.}~\bibnamefont{Raizen}},
  \bibinfo{author}{\bibfnamefont{H.}~\bibnamefont{Kimble}}, \bibnamefont{and}
  \bibinfo{author}{\bibfnamefont{P.}~\bibnamefont{Rice}},
  \bibinfo{journal}{Phys. Rev. A} \textbf{\bibinfo{volume}{40}},
  \bibinfo{pages}{5516} (\bibinfo{year}{1989}).

\bibitem[{\citenamefont{Cirac et~al.}(1997)\citenamefont{Cirac, Zoller, Kimble,
  and Mabuchi}}]{cirac1997quantum}
\bibinfo{author}{\bibfnamefont{J.}~\bibnamefont{Cirac}},
  \bibinfo{author}{\bibfnamefont{P.}~\bibnamefont{Zoller}},
  \bibinfo{author}{\bibfnamefont{H.}~\bibnamefont{Kimble}}, \bibnamefont{and}
  \bibinfo{author}{\bibfnamefont{H.}~\bibnamefont{Mabuchi}},
  \bibinfo{journal}{Phys. Rev. Lett.} \textbf{\bibinfo{volume}{78}},
  \bibinfo{pages}{3221} (\bibinfo{year}{1997}).

\bibitem[{khu(2008)}]{khurgin2008slow}
\emph{\bibinfo{title}{Slow light: Science and applications}}, vol.
  \bibinfo{volume}{140} (\bibinfo{publisher}{CRC press}, \bibinfo{year}{2008}).

\bibitem[{ote(2008)}]{otey2008completely}
\bibinfo{journal}{Journal of Lightwave Technology}
  \textbf{\bibinfo{volume}{26}}, \bibinfo{pages}{3784} (\bibinfo{year}{2008}).

\end{thebibliography}
\end{document}